\documentclass[11pt]{article}
\usepackage[utf8]{inputenc}
\pdfoutput=1
\usepackage{multirow}
\usepackage{amsfonts}
\usepackage{amsmath}
\usepackage{amssymb}
\usepackage{amsthm}
\usepackage{physics}
\usepackage{mathtools}
\usepackage{enumitem}
\usepackage{caption}
\usepackage[skip=10pt]{subcaption}
\usepackage[dvipsnames]{xcolor}
    \definecolor{darkgreen}{rgb}{0,0.5,0}
    \definecolor{darkblue}{rgb}{0,0,0.6}
    \definecolor{purple}{rgb}{0.4,.2,0.7}
\usepackage[margin = 2.5cm]{geometry}
\usepackage{graphicx}
\usepackage{tikz}
\usepackage{pgfplots}
\usepackage{pgfplotstable}
\usepackage{standalone}
\usepackage{hyperref}
\hypersetup{colorlinks = true, linkcolor = darkblue, citecolor = purple}
\usepackage{subcaption}
\usepackage{multirow}
\usepackage{hhline}
\usepackage{ytableau}
\usepackage{nccmath}
\usepackage{nicematrix}
\usepackage{simpler-wick}
\usepackage{tabularray}
\NiceMatrixOptions{cell-space-limits = 5pt}
\usepackage[noabbrev]{cleveref}
\usepackage{cite}
\usepackage{braket}

\usepackage{comment}

\definecolor{myblue1}{RGB}{33,113,181}
\definecolor{myblue2}{RGB}{158,202,225}
\definecolor{mygreen1}{RGB}{49,163,84}
\definecolor{mygreen2}{RGB}{161,217,155}
\definecolor{cycle1}{RGB}{128,177,211}
\definecolor{cycle2}{RGB}{253,180,98}
\definecolor{cycle3}{RGB}{179,222,105}
\definecolor{cycle4}{RGB}{251,128,114}
\definecolor{cycle5}{RGB}{190,186,218}
\definecolor{cycle6}{RGB}{141,211,199}

\newcommand{\ii}{\operatorname{i}}
\newcommand{\ee}{\operatorname{e}}
\newcommand{\col}{\operatorname{col}}

\newcommand{\len}{\operatorname{len}} 
\newcommand{\ad}{\operatorname{ad}}

\newcommand{\vspan}{\operatorname{span}}

\newcommand{\arctanh}{\operatorname{arctanh}}
\newcommand{\mycosh}{\operatorname{ch}}
\newcommand{\mysinh}{\operatorname{sh}}
\newcommand{\ch}{\operatorname{ch}}
\newcommand{\sh}{\operatorname{sh}}

\pgfplotsset{compat=newest}

\numberwithin{equation}{section}

\allowdisplaybreaks

\begin{document}




\thispagestyle{empty}
\begin{center}
    ~\vspace{5mm}

  \vskip 2cm 
  
   {\LARGE \bf 
       Thermofield Theory of Large {\bf \textit{N} Matrix Models} 
   }

   \vspace{0.5in}
     
   {\bf Antal Jevicki$^{a}$, Xianlong Liu$^{b,a}$, and Junjie Zheng$^{a}$}

    \vspace{0.5in}
    {$^{a}$\it Department of Physics, Brown University,\\
    182 Hope Street, Providence, RI 02912, United States
    }
  
    \vspace{0.2in}    
    {$^{a}$\it Brown Theoretical Physics Center, Brown University,\\ 
    340 Brook Street, Providence, RI 02912, United States}
                
    \vspace{0.2in}
    {$^{b}$\it Shanghai Center for Complex Physics, School of Physics and Astronomy,\\ 
    Shanghai Jiao Tong University, Shanghai 200240, China}

    \vspace{0.5in}

\end{center}

\vspace{0.5in}

\begin{abstract}

We develop analytical and numerical methods for the matrix thermofield in the large $N$ limit. Through the double collective representation on the Schwinger-Keldysh contour, it provides thermodynamical properties and finite temperature correlation functions, for large $N$ matrix quantum systems.

\end{abstract}

\vspace{1in}

\pagebreak


\setcounter{tocdepth}{3}
{\hypersetup{linkcolor=black}\tableofcontents}

\section{Introduction}
\label{sec:Introduction}

Matrix theory on the Schwinger-Keldysh contour is of direct relevance to the dual description \cite{Israel:1976ur,Maldacena:2001kr} of two-sided black holes. Even though the large $N$ limit (and the corresponding $1/N$ expansion) is expected to provide bulk properties of black hole space time (and Hilbert space), the limit has not been studied in concrete terms. The reason for this is the high non-perturbative complexity, involving summation of matrix model planar diagrams, and even more nontrivial evaluation of $1/N$ effects. Generally thermal properties of matrix models have been studied directly, through Monte-Carlo evaluations of \cite{Anagnostopoulos:2007fw,Hanada:2013rga} and very recently by thermal bootstrap methods in \cite{Cho:2024kxn,Cho:2024owx}.
Earlier studies involved variational \cite{Kabat:2000zv}, low
temperature \cite{Lin:2013jra} and large $D$ expansion \cite {Mandal:2009vz,Asano:2020yry} techniques.

\paragraph{}
In the present work, which is somewhat complementary, we present the construction of matrix thermofield at Large $N$. This construction builds on the insight gained in exact solution of related vector model problems given in \cite{Jevicki:2021ddf}. As in earlier studies, we consider the global gauging (of the U($N$) symmetry) which corresponds to the ungauged Hilbert space of the matrix system. Central to the construction that follows is the emergence of a continuous symmetry operating on the Schwinger-Keldysh contour. This allows the identification of the thermal solution, with the symmetry parameter being related to the temperature of the system. The collective background and other properties are found numerically using large $N$ optimization techniques that were successfully developed for ground state \cite{Koch:2021yeb,Mathaba:2023non}.

\paragraph{}
Thermodynamics can be studied in the Hamiltonian framework of the thermofield double (TFD) \cite{Takahasi:1974zn} formalism which we develop in this work. It involves the thermofield double state $\ket{0(\beta)}$ which has the property that it gives the thermal average for an arbitrary operator $\mathcal{O}$ as an expectation value
\begin{equation}
    \bra{0(\beta)} \mathcal{O} \ket{0(\beta)} = \frac{1}{Z(\beta)} \operatorname{Tr}\big(e^{-\beta H} \mathcal{O}\big) \, ,
\end{equation}
with $Z(\beta)$ the canonical partition function. This is possible through the introduction of an auxiliary (double) Hilbert space, representing a copy of the original one. 

\paragraph{}
This purification of the thermal state is not merely a mathematical trick, but also has significant physical relevance. In particular, the TFD state of the boundary CFTs is dual to the Hartle-Hawking state of an eternal black hole \cite{Maldacena:2001kr} and the Hilbert space built on it plays a role in the discussion of black hole information paradox. While the imaginary time dynamics is governed by the sum of two Hamiltonians ($\tilde{H}$ denotes the Hamiltonian of the doubled Hilbert space)
\begin{equation} \label{eq:H_plus}
    H_+ = H + \tilde{H} \, ,
\end{equation}
time evolution along the real part of the contour is generated by the thermofield Hamiltonian:
\begin{equation} \label{eq:H_hat}
    H_- = H - \tilde{H} \, .
\end{equation}
It will be this dynamics that we will be concerned with, with interest in developing methods for solving it at large $N$. We study this limit in the collective Hamiltonian \cite{Koch:2021yeb} representation. This approach provides not only the large $N$ values of thermal energy and free energy, but also correlation functions, spectrum, and representation of emergent Hilbert space. Several nontrivial features arise when attempting optimization of the thermofield problem. Among them are unboundedness of the Hamiltonian, the emergence of temperature and the uniqueness of the solution. Some of these were understood in our previous works on the subject, done in for vector type field theories which allowed for analytical solution. We will refer to these throughout this work \cite{Jevicki:2021ddf}.

\paragraph{}
The content of the paper is structured as follows: after reviewing the Hamiltonian loop space representation in \Cref{sec:collective_thermofield_Hamiltonian}, we concentrate on the free case in \Cref{sec:Free_theory} with emphasis on the emergent thermal symmetry that will be central to our work. In \Cref{sec:Interacting_theory} we present the numerical scheme with results for more general interacting theories. \Cref{appendix:loop_space_and_loop_functions} presents a detailed introduction to loop space representations, and \Cref{appendix:thermal_loop_values} presents the analytical computation of thermal average of loops in free theory.

\section{Collective thermofield Hamiltonian}
\label{sec:collective_thermofield_Hamiltonian}

For a general matrix quantum mechanical system, we are dealing with two Hamiltonians $H$ and $\tilde{H}$ 
\begin{align}
    H = \sum_i\frac{1}{2} \operatorname{tr}(\Pi_i^2) + V(\{M_i\}) \, , \\
    \tilde{H} = \sum_i\frac{1}{2} \operatorname{tr}(\tilde{\Pi}_i^2) + V(\{\tilde{M}_i\}) \, ,
\end{align}
where $\Pi_i$ ($\tilde{\Pi}_i$) denotes the canonical conjugate of $M_i$ ($\tilde{M}_i$). In Hilbert space, we have energy eigenstates of the form
\begin{equation}
    A_{i_1, j_1}^\dagger A_{i_2, j_2}^\dagger \cdots A_{i_n, j_n}^\dagger \ket{0} \, , \quad \tilde{A}_{i_1, j_1}^\dagger \tilde{A}_{i_2, j_2}^\dagger \cdots \tilde{A}_{i_n, j_n}^\dagger \ket{0} \, ,
\end{equation}
respectively. The thermofield double state, representing the vacuum of $H_-$ can be generated by $H_+$ as
\begin{equation}
    \ket{0(\beta)} = \frac{1}{\sqrt{Z(\beta)}} \ee^{-\beta H_+/4} \ket{I} \, ,
\end{equation}
where $\ket{I}$ is the (unnormalized) infinite temperature state, i.e., the maximally entangled state,
\begin{equation}
    \ket{I} = \sum_n \ket{n} \otimes \ket{\tilde{n}} \, .
\end{equation}
Explicitly, one has
\begin{equation}
    \ket{0(\beta)} = \frac{1}{\sqrt{Z(\beta)}} \sum_{\{i\}, \{j\}} \ee^{- \frac{1}{2} \beta E_{\{i\}, \{j\}}}A_{i_1, j_1}^\dagger A_{i_2, j_2}^\dagger \cdots A_{i_n, j_n}^\dagger \tilde{A}_{i_1, j_1}^\dagger \tilde{A}_{i_2, j_2}^\dagger \cdots \tilde{A}_{i_n, j_n}^\dagger \ket{0, 0} \, ,
\end{equation}
representing the thermal entangled state. This features a global U($N$) symmetry, which will allow the use of \textit{gauged} loops in the doubled space. The thermal state, however (and henceforth the model we consider in this work), corresponds to the \textit{ungauged} Hilbert space of the basic theory.

\paragraph{}
Let us clarify more precisely the two different gaugings: the gauge we consider above  can be termed as the \textit{diagonal} or \textit{global gauge}, with the single U($N$) gauge invariance
\begin{equation}
    U\{M_i\}U^\dagger\, , \quad  U\{\tilde{M}_i\}U^\dagger\, ,
\end{equation}
representing the total U$(N)$ gauge symmetry. The other possible gauging is known as the \textit{left-right gauge}, where the gauge symmetry of left and right copies are preserved separately, e.g., ${\rm{U}}(N)\times {\rm{U}}(\tilde{N})$ corresponding to respective ``singlet" subspaces. The former gauge group is smaller than the latter, and therefore more states survive under the diagonal gauge symmetry.

\paragraph{}
Continuing with the thermofield dynamics of the global gauged theory, we build it in terms of the associated single-trace loop variables $\phi(C) \equiv \tr(C) / N^{\len(C)/2 + 1}$, with $C$ a word built from the alphabet $\{M, \tilde{M}\}$. The collective construction of the (real-time) Hamiltonian $H_-$ features the loop joining and loop splitting functions 
\begin{align}
    \Omega_-(C_1, C_2) & \equiv \Omega_{11}(C_1, C_2) - \Omega_{22}(C_1, C_2) \, , \\ 
    \omega_-(C) & \equiv \omega_{11}(C) - \omega_{22}(C) \, .
\end{align}
We refer the reader to \Cref{appendix:loop_space_and_loop_functions} for a detailed review of the loop space and loop space functions such as $\Omega_{ab}(C_1, C_2)$ and $\omega_{ab}(C)$ on the right hand sides. Here the subscript $\{1, 2\}$ of $\Omega$ and $\omega$ refers to copies of $\{M_1, M_2\} \equiv \{M, \tilde{M}\}$ (we adopt these two notations interchangeably). The loop space thermofield Hamiltonian then takes the form
\begin{equation}
    \hat{H}_- = \frac{1}{2 N^2} P(C_1) \Omega_-(C_1, C_2) P(C_2) + N^2 \hat{V}_-[\phi(C)] \, , 
\end{equation}
with the nonlinear \emph{thermofield collective potential} $\hat{V}_-[\phi]$ 
\begin{equation}
    \hat{V}_-[\phi] = \frac{1}{8} \omega_-(C_1) \Omega_-^{-1}(C_1, C_2) \omega_-(C_2) + V_-[\phi(C)] \, .
\end{equation}
In the above the summations over $C_1$ and $C_2$ are assumed. $V_-[\phi]$ denotes the difference of the original potential term represented in loop space: $V_- = N^{-2} \tr\big(V(M_1) - V(M_2)\big)|_{\phi}$. For example, take $V(M) = M^2 / 2$, then we have $V_- = \phi(M_1^2) / 2 - \phi(M_2^2) / 2$, with $\phi$ the associated loops. 

\paragraph{}
In addition, we also have ``positive semi-definite" joining matrix $\Omega_+ \equiv \Omega_{11} + \Omega_{22}$ which will play two roles. First it enters the imaginary time evolution Hamiltonian $H_+$, and then due to Cauchy-Schwarz inequalities of the matrix products we have that  $\Omega_{+}$ is positive semi-definite: $\Omega_{+} \succeq 0$. The loop joining matrix $\Omega_+$ therefore generates the  positivity constraints, also central in the ``bootstrap'' method \cite{Anderson:2016rcw,Lin:2020mme,Han:2020bkb,Kazakov:2021lel,Li:2024ggr,Cho:2024kxn}. Similar to the earlier work \cite{Koch:2021yeb}, the large $N$ thermal backgrounds will be achieved as solutions of the saddle point equation that satisfy such constraints
\begin{equation}
    \label{eq:cstr}
    \begin{dcases}
        \Omega_+(C, C^\prime) \pdv{\hat{V}_-[\phi]}{\phi(C^\prime)} = 0 \, , \\
        \Omega_{+} = \Omega_{11} + \Omega_{22} \succeq 0 \, .
    \end{dcases}
\end{equation}
Instead of using the first equation of \eqref{eq:cstr}, one can alternatively use the ``dual" equation
\begin{equation}
    \label{eq:dual}
    \Omega_-(C, C^\prime) \pdv{\hat{V}_+[\phi]}{\phi(C^\prime)} = 0 \, ,
\end{equation}
where $\hat{V}_{+}$ denotes the collective potential for $H_+$. These two sets of stationary equations are not unrelated; they in fact imply each other, as a consequence of the $G$ symmetry, e.g. $[G, H_-] = 0$ that will be established below.

\paragraph{}
As the example of the O($N$) vector model already reveals \cite{Jevicki:2015sla,Jevicki:2021ddf}, one expects a class of infinitely many solutions of this equation. Each solution represents the thermal background at a specific temperature, and satisfies $\langle \hat{V}_- \rangle = 0$.  As was the case at zero temperature, the thermal background can be obtained by the numerical optimization scheme presented in \cite{Koch:2021yeb} that utilizes ``master field variables", turning the constrained minimization problem into an unconstrained minimization problem. The difficulty here is that, in contrast to the cases in \cite{Koch:2021yeb} where all collective potentials have a lower bound, the thermofield collective potential $\hat{V}_-$ is unbounded from below. On the other hand, at the large $N$ background, the gradient of the collective potential with respect to the master fields should be 0. Thus the problem is approachable if one can perform the minimization of the square of the gradient of $\hat{V}_-$ with respect to master fields. Minimizing the square of the gradient requires significantly more computational resources, which is more challenging than the original optimization scheme in \cite{Koch:2021yeb}. Nevertheless, as we show, the numerical optimization scheme is still feasible for relatively small truncations of the loop space. We will present a detailed numerical framework with results for interacting theories in \Cref{sec:Interacting_theory}. In addition, for the free theory case we give a much simpler optimization problem in \Cref{sec:Free_theory}.

\section{Free theory}
\label{sec:Free_theory}

Study of the free Gaussian matrix theory is already of certain relevance. In Yang-Mills and vector model cases it features a nontrivial Hagedorn confinement-deconfinement phase transition \cite{Sundborg:1999ue,Aharony:2003sx,Shenker:2011zf} in the singlet sector of the theories. Furthermore, at large $N$ in the collective framework it represents a nonlinear theory, the nonlinearity being generated by the collective potential. A numerical investigation and the numerical solution is therefore as nontrivial as in the presence of interaction. Furthermore, as it will be discussed, the theory features a symmetry, which in this case can be fully understood as being associated with thermal Bogoliubov transformations. This symmetry will be used for exact evaluation of thermal correlators. It will also play a role in the numerical optimization.

\subsection{Symmetry transformations}

The Hamiltonian governing the time evolution is
\begin{equation} \label{eq:H_hat_2}
    H_- = H_1 - H_2 = 
    \frac{1}{2} \operatorname{tr}[ (\Pi_1^2 + M_1^2) - (\Pi_2^2 + M_2^2) ] \, .
\end{equation}
In free theory case, one also has the following $G$ symmetry operator 
\begin{equation} \label{eq:G}
    G = \theta(\beta)  \tr(M_{1} \Pi_{2} + M_{2} \Pi_{1}) 
    \, ,
    \quad \text{with} \quad 
    \theta(\beta) = \arctanh(\ee^{-\beta / 2}) \, 
\end{equation}
generating thermal Bogoliubov transformations and the thermofield state: 
\begin{equation}
    \ket{0(\beta)} = \ee^{- \ii G} \ket{0}_{1} \otimes \ket{0}_{2} \, .
\end{equation}
Using commutation relations
\begin{equation}
    [(M_i)^{a}{}_{b}, (\Pi_j)^{c}{}_{d}] = \ii \delta_{ij} \delta^{a}{}_{d} \delta^{c}{}_{b} \, , \quad i, j = 1, 2 \, ,
\end{equation}
one can check that the aforementioned $G$ operator is a conserved quantity, i.e., it commutes with the Hamiltonian $H_-$.

\paragraph{}
We see that the TFD state is obtained by a unitary transformation from the ground state induced by the $G$ operator, representing a Bogoliubov transformation. To compute thermal quantities, one can just compute the expectation values of the corresponding operators in $\ket{0(\beta)}$. This resembles the Schr\"{o}dinger picture in quantum mechanics. On the other hand, we can adopt the ``Heisenberg'' picture, namely we fix the states, and transform operators $\mathcal{O}$ as\footnote{
    Note that here the signs are opposite to the transformations in \cite{Takahasi:1974zn}, hence we put a minus sign in $\mathcal{O}_{-\theta}$.
}
\begin{equation}
    \mathcal{O}_{- \theta} = \ee^{\ii G} \mathcal{O} \ee^{-\ii G} \, .
\end{equation}
Similarly, to calculate thermal quantities, one can compute the expectation value of $\mathcal{O}_{\beta}$ with respect to the ground state. That is,
\begin{equation} \label{eq:Bogoliubov_property}
    \bra{0(\beta)} \mathcal{O} \ket{0(\beta)} = \bra{0} \mathcal{O}_{- \theta} \ket{0} \, ,
\end{equation}
where $\ket{0} \equiv \ket{0}_1 \otimes \ket{0}_2$ denotes the ground state. This relation, although seemingly trivial, will play a fundamental role in our analytical calculations. We emphasize that this is not the real Heisenberg picture, since the transformation is induced by $G$ with a parameter $\theta(\beta) $, instead of the time evolution of operators based on the Hamiltonian.


\paragraph{}
We then move on to consider expectation values of loop variables at finite temperature, which can be computed directly from $G$-transformations of loops:\footnote{
  Here the superscript $a$ in $\phi^a$ represents ``loop index'', and is in one-to-one correspondence with the loop word $C$ to simplify the notation. One may refer to \Cref{tab:loopinfo2} for the correspondence of the first 16 loops. Note that here we use $\ad_{\ii G}$ instead of $\ad_{- \ii G}$ used in \cite{Takahasi:1974zn}.
}
\begin{equation} \label{eq:G_f_transform}
  \phi^{a}_{- f} \equiv \ee^{\ad_{\ii G_f}}(\phi^a) 
  = \sum_{n=0}^{\infty} \frac{1}{n!} \ad_{\ii G_f}^{n}(\phi^a) \, ,
\end{equation}
with $\ad_{\ii G_f} = \ii [G_f, \phi^{a}]$. For generality, we use $G_f$ defined as
\begin{equation}
  G_f = f(\beta) \tr(M_{1} \Pi_{2} + M_{2} \Pi_{1}) \, ,
\end{equation}
with $f(\beta)$ an arbitrary smooth function of $\beta$. All such $G_f$ operators commute with the thermofield Hamiltonian $H_-$, and hence represents a symmetry \cite{Jevicki:2021ddf}. For thermal equilibrium states at finite temperature, one identifies $f(\beta) = \theta(\beta)$. Due to the definition of $G_f$, we see that $G_f$-transformations preserve the length of a loop.\footnote{
  This is no longer true in interacting theories, since $G_f$ will then consist of higher order terms.
}
In other words, let $V_{\ell}$ denote the loop subspace spanned by all independent loops of length $\ell$, $G_f$ induces an isomorphism on $V_{\ell}$. Then for a loop $\phi^{a} \in V_{\ell}$, we must have
\begin{equation}
  \phi^{a}_{-f} = \ee^{\ad_{\ii G_f}}(\phi^{a}) \in V_{\ell} \, .
\end{equation}
Thus, in the case of free theory, the loop $\phi^{a}_{-f}$ is a linear combination of the basis of $V_{\ell}$, with linear coefficients some functions of $f(\beta)$:\footnote{
  In \cref{eq:W_mat}, to be more precise in the notations, the indices of the matrix $W$ should be $W^{a - d_{\ell}}{}_{b - d_{\ell}}$, with $d_{\ell} = \sum_{m=1}^{\ell-1} \dim(V_{m})$, such that they range from $1$ to $\dim(V_{\ell})$. In the following discussions we adopt the same notation, which is simple and apparent.
}
\begin{equation} \label{eq:W_mat}
  \phi^{a}_{-f} = \sum_{\phi^{b} \in V_{\ell}} W_{\ell}{}^{a}{}_{b}[f(\beta)] \phi^{b} \in V_{\ell} \, .
\end{equation}
Here $W_{\ell}$ is a matrix of size $\dim(V_{\ell}) \times \dim(V_{\ell})$. This observation provides us a very simple and straightforward strategy to compute $G_f$-transformations: we simply compute $W_{\ell}$ and then multiply it with the loop values at zero temperature. To compute $W_{\ell}$ we can first expand the linear transformation matrix $W_{\ell}$ \cref{eq:W_mat} into its Taylor series:
\begin{equation}
  \phi_{-f}^{a} = \sum_{\phi^{b} \in V_{\ell}} \sum_{n=0}^{\infty} \frac{1}{n!} w_{\ell, n}{}^{a}{}_{b}[f] \, \phi^{b} \, .
\end{equation}
Comparing this with the $G_f$-transformation formula \cref{eq:G_f_transform} we see that
\begin{equation}
  \sum_{b} w_{\ell, n}{}^{a}{}_{b}[f] \,  \phi^{b} = \ad_{\ii G_f}^{n}(\phi^{a}) \, .
\end{equation}
From this we have $w_{\ell, n} = (w_{\ell, 1})^{n}$. Thus, all we need to do is to compute $w_{\ell, 1}$, and then exponentiate it to obtain $W_{\ell}$, i.e.
\begin{equation}
  W_{\ell} = \exp(w_{\ell}) \, , \qquad w_{\ell} \equiv w_{\ell, 1} \, .
\end{equation}

\paragraph{}
Due to the simple form of $G_f$ in free theory case, there are several important properties for $w_{\ell}$:
\begin{enumerate}[label=(\roman*)]
  \item When adding its matrix elements along an arbitrary row, the result should be $\ell \, f$:
  \begin{equation}
      \sum_{b = 1}^{\dim(V_{\ell})} w_{\ell}{}^{a}{}_{b} = \ell \, f \, , \qquad \forall a = 1 , \dots , \dim(V_{\ell}) \, .
  \end{equation}
  \item Consequently, when summing all its matrix elements, the results should be
  \begin{equation}
      \sum_{a, b = 1}^{\dim(V_{l})} w_{\ell}{}^{a}{}_{b} = \dim(V_{\ell}) \, \ell \, f \, .
  \end{equation}
  \item We also note that the diagonal elements of $w_{\ell}$ are always 0:
  \begin{equation}
      w_{\ell}{}^{a}{}_{a} = 0 \, \, \quad (\text{no summation}).
  \end{equation}
  This simply means $\phi^{a}$ itself is absent in $\ad_{\ii G_f}(\phi^{a})$. 
\end{enumerate}
As a result of the last property, $w_{\ell}$ is always traceless: $\tr(w_{\ell}) = 0$. Thus, the determinant of $W_{\ell}$ is always $1$ for all $\ell \in \mathbb{N}$:
\begin{equation}
  \det(W_{\ell}) = \exp(\tr(w_{\ell})) = 1 \, , 
  \quad \Rightarrow \quad
  W_{\ell} \in \operatorname{SL}(\dim(V_{\ell}), \mathbb{R}) \, .
\end{equation}
Below we will present concrete examples to illustrate these properties and compute thermal averages of loops.

\paragraph{Transformations in $V_{1}$.}
First we calculate $G_f$-transformation for loops in $V_{1}$. This case can be computed directly without using matrix exponentiation of $w_{\ell}$. For $\phi^{1} = \tr(M_1) / N^{3/2}$ we have
\begin{equation}
    \ad_{\ii G_f}^{2k}(\phi^1) = f^{2k} \phi^1 \, , 
    \qquad
    \ad_{\ii G_f}^{2k+1}(\phi^1) = f^{2k+1} \phi^2 \, ,
    \qquad
    k \in \mathbb{N} \, .
\end{equation}
Thus we find that
\begin{equation}
    \phi^{1}_{-f} = \cosh(f) \phi^{1} + \sinh(f) \phi^{2} \, .
\end{equation}
Similarly for $\phi^{2} = \tr(M_2) / N^{3/2}$ we have
\begin{equation}
    \phi^{2}_{-f} = \sinh(f) \phi^{1} + \cosh(f) \phi^{2} \, .
\end{equation}
Putting these into matrix form, we find the linear transformation matrix $W_{1}$ as
\begin{equation}
    \begin{bmatrix}
        \phi^{1}_{-f} \\[5pt] \phi^{2}_{-f}
    \end{bmatrix}
    =
    \begin{bmatrix}
        \cosh(f) & \sinh(f) \\[5pt]
        \sinh(f) & \cosh(f)
    \end{bmatrix}
    \begin{bmatrix}
        \phi^{1} \\[5pt] \phi^{2}
    \end{bmatrix} \, .
\end{equation}
We can easily see that our strategy provided above agrees precisely with the direct computation:
\begin{equation}
    w_{1} = \begin{bmatrix}
        0 & 1 \\
        1 & 0
    \end{bmatrix} \, ,
    \quad \Rightarrow \quad
    W_{1} = \exp(w_{1}) = 
    \begin{bmatrix}
        \cosh(f) & \sinh(f) \\
        \sinh(f) & \cosh(f)
    \end{bmatrix} \, .
\end{equation}
Since in the free theory case $\phi^{1} = \phi^{2} = 0$ at the zero temperature, this relation tells us that at finite temperature we also have $\phi^{1}_{\beta} = \phi^{2}_{\beta} = 0$.\footnote{
    To avoid notational clutters, when we discuss the expectation values, $\phi^{a} \equiv \langle 0 | \phi^{a} | 0 \rangle$ refers to the expectation values at ground state, and $\phi^{a}_{\beta} \equiv \langle 0(\beta) | \phi^{a} | 0(\beta) \rangle$ refers to the thermal averages.
} 
Thus we conclude that in the free theory,
\begin{equation}
    \phi^{1}_{\beta} = \phi^{2}_{\beta} = 0 \, , 
    \quad
    \forall \beta \, .
\end{equation}

\paragraph{Transformations in $V_{2}$.}
We next consider the $G_f$-transformation for loops in 
\[ 
    V_{2} = \vspan\{\phi^3, \phi^4, \phi^5\} = \vspan\{\tr(M_1^2), \tr(M_1 M_2), \tr(M_2^2)\} \, . 
\] 
In this case we have
\begin{equation}
    \ad_{\ii G_f}
    \begin{bmatrix}
        \phi^{3} \\ \phi^{4} \\ \phi^{5}
    \end{bmatrix}
    =
    f
    \begin{bmatrix}
        0 & 2 & 0 \\
        1 & 0 & 1 \\
        0 & 2 & 0
    \end{bmatrix}
    \begin{bmatrix}
        \phi^{3} \\ \phi^{4} \\ \phi^{5}
    \end{bmatrix} \, .
\end{equation}
%
With these commutation relations we find that
\begin{equation}
    \begin{bmatrix}
        \phi^{3}_{-f} \\[5pt] \phi^{4}_{-f} \\[5pt] \phi^{5}_{-f}
    \end{bmatrix}
    =
    \begin{bmatrix}
        \cosh ^2(f) & \sinh (2 f) & \sinh ^2(f) \\[5pt]
        \frac{1}{2} \sinh (2f) & \cosh (2 f) & \frac{1}{2} \sinh (2f) \\[5pt]
        \sinh ^2(f) & \sinh (2 f) & \cosh ^2(f) \\[5pt]
    \end{bmatrix}
    \begin{bmatrix}
        \phi^{3}{}_{ } \\[5pt] \phi^{4}{}_{ } \\[5pt] \phi^{5}{}_{ }
    \end{bmatrix} \, .
\end{equation}
Since at zero temperature we have $\phi^{3} = \phi^{5} = \frac{1}{2}$, and $\phi^{4} = 0$, at finite temperature we have
\begin{equation}
    \phi^{3}_{\beta} = \phi^{5}_{\beta} = \frac{1}{2} \cosh(2\theta(\beta)) \, ,
    \qquad
    \phi^{4}_{\beta} = \frac{1}{2} \sinh(2\theta(\beta)) \, .
\end{equation}
With these results we also find an operator which commutes with $G$, and hence is constant for all temperature:
\begin{equation} \label{eq:V2_invariant}
    4 \left[\phi^{3}_{\beta} \phi^{5}_{\beta} - (\phi^{4}_{\beta})^{2}\right] = 1 \, ,
    \quad
    \forall \beta \, .
\end{equation}
This relation can be used to check if we perform well in the numerical optimizations framework presented in the following subsection.

\paragraph{}
The above examples might be too trivial because $V_{2}$ can be regarded as the ``vector model'' sector in the loop space of matrices. In fact, these results are exactly what we obtained in the O($N$) vector models \cite{Jevicki:2021ddf}. It is therefore more interesting to investigate $V_{\ell}$ with larger $\ell$'s. 

\paragraph{Transformations in $V_{3}$.}
Let us now study the $G_f$-transformation of 
\[
    V_{3} = \vspan{\{\phi^{6}, \phi^{7}, \phi^{8}, \phi^{9}\}}
     = \vspan{\{\tr(M_1^3), \tr(M_1^2 M_2), \tr(M_1 M_2^2), \tr(M_2^3) \}} \, . 
\] 
We have $w_{3}$ as
\begin{equation}
    \ad_{\ii G_f} 
    \begin{bmatrix}
        \phi^{6} \\ \phi^{7} \\ \phi^{8} \\ \phi^{9}
    \end{bmatrix}
    =
    f
    \begin{bmatrix}
        0 & 3 & 0 & 0 \\
        1 & 0 & 2 & 0 \\
        0 & 2 & 0 & 1 \\
        0 & 0 & 3 & 0 \\
    \end{bmatrix}
    \begin{bmatrix}
        \phi^{6} \\ \phi^{7} \\ \phi^{8} \\ \phi^{9}
    \end{bmatrix} \, .
\end{equation}
The matrix exponentiation then gives\footnote{For notational simplicity we use $\ch(f) \equiv \cosh(f)$ and $\sh(f) \equiv \sinh(f)$.}
\begin{equation}
    W_{3}[f] = \frac{1}{4}
    \begin{bmatrix}
        3 \mycosh(f) + \mycosh(3f) & 3 \mysinh(f) + 3 \mysinh(3f) & -3 \mycosh(f) + 3 \mycosh(3f) & -3 \mysinh(f) + \mysinh(3f) \\[5pt]
        \mysinh(f) + \mysinh(3f) & \mycosh(f) + 3 \mycosh(3f) & - \mysinh(f) + 3 \mysinh(3f) & - \mycosh(f) + \mycosh(3f) \\[5pt]
        - \mycosh(f) + \mycosh(3f) & - \mysinh(f) + 3 \mysinh(3f) & \mycosh(f) + 3 \mycosh(3f) & \mysinh(f) + \mysinh(3f) \\[5pt]
        -3 \mysinh(f) + \mysinh(3f) & -3 \mycosh(f) + 3 \mycosh(3f) & 3 \mysinh(f) + 3 \mysinh(3f) & 3 \mycosh(f) + \mycosh(3f)
    \end{bmatrix} \, .
\end{equation}
One can verify that the determinant of $W_{3}$ is always 1:
\begin{equation}
    \det(W_{3}[f]) = 1 \, , \quad \forall f(\beta) \, .
\end{equation}
Again, since all loops in $V_{3}$ are 0 at zero temperature, they remain to be 0 at finite temperature:
\begin{equation}
    \phi_{\beta}^{a} = 0 \, , \quad \forall \phi^{a} \in V_{3} \, .
\end{equation}

\paragraph{Transformations in $V_{4}$.}
We then compute the $G_f$-transformation of 
\begin{align*}
    V_{4} & = \vspan(\{\phi^{10}, \dots , \phi^{15}\}) \\
    & = \vspan\{\tr(M_1^4), \tr(M_1^3 M_2), \tr(M_1^2 M_2^2), \tr(M_1 M_2 M_1 M_2), \tr(M_1 M_2^3), \tr(M_2^4) \} \, .
\end{align*}
We have $w_{4}$ as
\begin{equation}
    \ad_{\ii G_f} \begin{bmatrix}
        \phi^{10} \\ \phi^{11} \\ \phi^{12} \\ \phi^{13} \\ \phi^{14} \\ \phi^{15} 
    \end{bmatrix}
    =
    f
    \begin{bmatrix}
        0 & 4 & 0 & 0 & 0 & 0 \\
        1 & 0 & 2 & 1 & 0 & 0 \\
        0 & 2 & 0 & 0 & 2 & 0 \\
        0 & 2 & 0 & 0 & 2 & 0 \\
        0 & 0 & 2 & 1 & 0 & 1 \\
        0 & 0 & 0 & 0 & 4 & 0
    \end{bmatrix}
    \begin{bmatrix}
        \phi^{10} \\ \phi^{11} \\ \phi^{12} \\ \phi^{13} \\ \phi^{14} \\ \phi^{15} 
    \end{bmatrix} \, .
\end{equation}
Upon matrix exponentiation, we obtain 
\begin{equation}\medmath{
\begin{aligned}
    & W_{4} = \\
    & \begin{bmatrix}
        \ch ^4(f) & 4 \sh (f) \ch^3(f) & \sh ^2(2 f) & 2 \sh ^2(f) \ch ^2(f) & 4 \sh ^3(f) \ch (f) & \sh ^4(f) \\[5pt]
 \sh (f) \ch ^3(f) & \frac{1}{2} (\ch (2 f)+\ch (4 f)) & \frac{1}{2} \sh (4 f) & \frac{1}{4} \sh (4 f) & \sh ^2(f) (2 \ch
   (2 f)+1) & \sh ^3(f) \ch (f) \\[5pt]
 \sh ^2(f) \ch ^2(f) & \frac{1}{2} \sh (4 f) & \ch ^2(2 f) & 2 \sh ^2(f) \ch ^2(f) & \frac{1}{2} \sh (4 f) & \sh ^2(f)
   \ch ^2(f) \\[5pt]
 \sh ^2(f) \ch ^2(f) & \frac{1}{2} \sh (4 f) & \sh ^2(2 f) & \frac{1}{4} (\ch (4 f)+3) & \frac{1}{2} \sh (4 f) & \sh ^2(f)
   \ch ^2(f) \\[5pt]
 \sh ^3(f) \ch (f) & \sh ^2(f) (2 \ch (2 f)+1) & \frac{1}{2} \sh (4 f) & \frac{1}{4} \sh (4 f) & \frac{1}{2} (\ch (2 f)+\ch
   (4 f)) & \sh (f) \ch ^3(f) \\[5pt]
 \sh ^4(f) & 4 \sh ^3(f) \ch (f) & \sh ^2(2 f) & 2 \sh ^2(f) \ch ^2(f) & 4 \sh (f) \ch ^3(f) & \ch ^4(f)
    \end{bmatrix} \, . \\[5pt]
\end{aligned}}
\end{equation}
In this case we also have
\begin{equation}
    \det(W_{4}[f]) = 1 \, , \quad \forall f(\beta) \, .
\end{equation}
With the $W$ matrix we can then compute the loop variables at finite temperature in $V_{4}$. At zero temperature we have
\begin{equation}
    \phi^{10} = \phi^{15} = \frac{1}{2} \, , \qquad
    \phi^{12} = \frac{1}{4} \, , \qquad
    \phi^{11} = \phi^{13} = \phi^{14} = 0 \, .
\end{equation}
Thus at finite temperature we have
\begin{align}
    & \phi^{10}_{\beta} = \phi^{15}_{\beta} = \frac{1}{2} \ch^2(2 \theta(\beta)) \, , 
    & \phi^{13}_{\beta} = \frac{1}{2} \sh^2(2\theta(\beta)) \, , \\
    & \phi^{11}_{\beta} = \phi^{14}_{\beta} = \frac{1}{4} \sh(4 \theta(\beta)) \, ,
    & \phi^{12}_{\beta} = \frac{1}{4} \ch(4 \theta(\beta)) \, .
\end{align}
From these results we see that in free theory there are also sub-invariants in $V_{4}$, namely,
\begin{align} \label{eq:V4_sub_invariants_1}
    \phi^{10}_{\beta} + \phi^{15}_{\beta} - 2 \phi^{13}_{\beta} & = 1 \, , \\
    \label{eq:V4_sub_invariants_2}
    16 [( \phi^{12}_{\beta})^2 - \phi^{11}_{\beta} \phi^{14}_{\beta}] & = 1 \, .
\end{align}

\paragraph{}
The dimension of $V_{\ell}$ increases rapidly when $\ell$ increases. It therefore becomes cumbersome for computing the $G_f$-transformation matrix $W$ for larger $\ell$ by hand. One can directly use computer programming to solve it, just as how we obtain the loop joining and splitting in the loop space with master field optimization \cite{Koch:2021yeb}. It might also be interesting to investigate the sub-invariants in $V_{\ell}$, such as what we find in $V_{4}$ \cref{eq:V4_sub_invariants_1,eq:V4_sub_invariants_2}. These relations should be related to the additional symmetries in $V_{\ell}$. In \Cref{tab:loopinfo2} we present all the loops in $V_{\ell}$ up to $\ell = 4$ and their expectation values at finite temperature in free theory. All higher loops, and loops in multi-matrix systems, can be computed in the same way. In \Cref{appendix:thermal_loop_values} we present nonzero higher loop values in both zero temperature and finite temperature, where we also present a simple strategy to compute the loops values at zero temperature in free theory.

\renewcommand{\arraystretch}{1.2}
\begin{table}[htb!]
    \begin{center}
    \begin{tabular}{|c|c|c|c|c|}
        \hhline{|=|=|=|=|=|}
        loop subspace & loop index & loop word & $\bra{0} \phi^{a} \ket{0}$ & $\bra{0(\beta)} \phi^{a} \ket{0(\beta)}$ \\
        \hline
        $V_{0}$ &  0          & []           & 1 & 1 \\
        \hline
        \multirow{2}{*}{$V_{1}$} &  1          & [1]    & 0      & 0 \\
        & 2          & [2]          & 0 & 0 \\
        \hline
        \multirow{3}{*}{$V_{2}$} & 3          & [11]    & $\frac{1}{2}$  & $\frac{1}{2} \cosh(2 \theta_{\beta})$ \\
        & 4          & [12]       & 0 & $\frac{1}{2} \sinh(2 \theta_{\beta})$ \\
        & 5          & [22]       & $\frac{1}{2}$ & $\frac{1}{2} \cosh(2 \theta_{\beta})$ \\
        \hline
        \multirow{4}{*}{$V_{3}$} & 6          & [111]   & 0  & 0 \\
        & 7          & [112]    & 0 & 0 \\
        & 8          & [122]    & 0 & 0 \\
        & 9          & [222]    & 0 & 0 \\
        \hline
        \multirow{6}{*}{$V_{4}$} & 10         & [1111] & $\frac{1}{2}$  & $\frac{1}{2} \cosh^2(2 \theta_{\beta})$ \\
        & 11         & [1112] & 0 & $\frac{1}{4} \sinh(4 \theta_{\beta})$ \\
        & 12         & [1122] & $\frac{1}{4}$ & $\frac{1}{4} \cosh(4 \theta_{\beta})$ \\
        & 13         & [1212] & 0 & $\frac{1}{2} \sinh^2(2 \theta_{\beta})$ \\
        & 14         & [1222] & 0 & $\frac{1}{4} \sinh(4 \theta_{\beta})$ \\
        & 15         & [2222] & $\frac{1}{2}$ & $\frac{1}{2} \cosh^2(2 \theta_{\beta})$ \\
        \hline
    \end{tabular}
     \caption{The first 16 loops and their expectation values in free theory}
    \label{tab:loopinfo2}
    \end{center}
    \end{table}

\paragraph{}
We find that the spectrum for the order 1 fluctuation $\hat{H}_-^{(2)}$ can be schematically written as a union of loop subspaces:
\begin{equation}
    \{\hat{\varepsilon}\} = \bigcup_{\ell = 1}^{\infty} \{\hat{\varepsilon}\}_{\ell} \, ,
\end{equation}
with 
\begin{equation}
    \{ \hat{\varepsilon} \}_{\ell} = \{ \underbrace{0, \dots, 0}_{d_{\ell, 0}} , \underbrace{1, \dots, 1}_{d_{\ell, 1}} , \underbrace{2, \dots, 2}_{d_{\ell, 2}} \dots, \underbrace{\ell-1, \dots , \ell -1}_{d_{\ell, \ell - 1}} , \underbrace{\ell, \ell}_{2}\} \, .
\end{equation}
The degeneracy at level $n$ is denoted $d_{\ell, n}$, and is given by numbers of loops in the basis of $V_{\ell}$ whose number of $M_1$ minus number of $M_2$ equals $\pm n$. For example, for the basis of $V_4$ we have two loops that have the same number of $M_1$ and $M_2$, namely $\phi(M_1^2 M_2^2)$ and $\phi(M_1 M_2 M_1 M_2)$. Thus we have two zero modes in $\{\hat{\varepsilon}\}_4$:
\begin{equation}
    \{\hat{\varepsilon}\}_4 = \{0, 0, 2, 2, 4, 4\} \, .
\end{equation}
Each zero eigenvalue corresponds to a symmetry operator. For example, $\{\hat{\varepsilon}\}_2 = \{0, 2, 2\}$, and the zero eigenvalue corresponds to the symmetry condition that
\begin{equation}
    \langle \phi(M_1^2) \rangle_{\beta} \langle \phi(M_2^2) \rangle_{\beta} - \langle \phi(M_1 M_2)^2 \rangle_{\beta} = \frac{1}{4} \, , \quad \forall \beta \, ,
\end{equation}
and the operator $\phi(M_1^2) \phi(M_2^2) - \phi(M_1 M_2)$ commutes both with $H_-$ and $G$. From \cref{eq:dim_Vl_generating_function} we find that the degeneracy pattern is given by
\begin{equation}
    d_{\ell, m} = \sum_{\ell \mid d} \frac{\varphi(d)}{\ell} \sum_{k=0}^{\ell/d} \delta_{m, |2kd-\ell|} \binom{\ell/d}{k} \, ,
\end{equation}
where $\varphi(d)$ is Euler's totient function.

\subsection{Numerical framework}

In this subsection we present the large $N$ numerical framework, which can be tested for evaluation of the above free thermal averages. We would like to find a quantity which is suitable for minimization at finite temperature. It turns out that the average energy operator is one such quantity:
\begin{equation} \label{eq:E}
    \mathcal{E} = \frac{1}{2} (H_1 + H_2) 
    = \frac{1}{4} \operatorname{tr}(\Pi_{1}^2 + \Pi_{2}^2 + M_{1}^2 + M_{2}^2) \, .
\end{equation}
The TFD state is related to the maximally entangled state $\ket{I}$ by a non-unitary transformation induced by it: 
\begin{equation}
    \ket{0(\beta)} = \frac{1}{\sqrt{Z(\beta)}} \ee^{- \beta \mathcal{E} / 2} \ket{I} \, .
\end{equation}
This operator also commutes with $H_-$. However, it does not commute with $G$, under whose transformation becomes
\begin{equation} \label{eq:E_beta}
    \mathcal{E}_{\theta} = \ee^{- \ii G} \mathcal{E} \ee^{\ii G}
    = \frac{1}{2} \left[
        \cosh(2 \theta(\beta)) \mathcal{E}
        + \sinh(2 \theta(\beta)) \operatorname{tr}(\Pi_1 \Pi_2 - M_1 M_2) 
    \right] \, .
\end{equation}
Now let us explain why $\mathcal{E}_{\theta}$ is the quantity to be minimized. According to \cref{eq:Bogoliubov_property}, the expectation value of $\mathcal{E}_{\theta}$ with respect to the TFD state $\ket{0(\beta)}$ is the ground state energy. Equivalently, $\ket{0(\beta)}$ is the ground state of $\mathcal{E}_{\theta}$. Therefore, when minimizing (the  collective potential associated to)~$\mathcal{E}_{\theta}$, we will obtain the TFD state $\ket{0(\beta)}$, with loop values their thermal averages, and the minimum the ground state energy.

\paragraph{}
To apply the numerical optimization framework established in \cite{Jevicki:1983hb,Koch:2021yeb}, we need to first compute the loop space representation. Essentially this involves the rewriting of the kinetic terms. In the average energy operator \eqref{eq:E_beta} there are two kinds of kinetic terms, namely $\tr \big( \Pi_1^2 + \Pi_2^2 \big)$ and $\tr \big( \Pi_1 \Pi_2 \big)$. We first study the collective representation of the first one. Using the commutation relation we can write
\begin{equation}
    \tr(\Pi_1^2 + \Pi_2^2) = - \tr(\pdv[2]{}{M_1}{M_1} +  \pdv[2]{}{M_2}{M_2}) \, .
\end{equation}
The corresponding loop joining $\Omega_{+}(C_1, C_2)$ and the loop splitting $\omega_{+}(C)$ are then
\begin{align}
    \Omega_{+} & = \Omega_{11} + \Omega_{22} \, , \\
    \omega_{+} & = \omega_{11} + \omega_{22} \, ,
\end{align}
where $\Omega_{ij}$ is defined \cref{eq:Omega_ij} in and $\omega_{ij}$ is defined in \cref{eq:omega_ij}. The first kinetic term in the collective field representation then is
%
\begin{equation}
    \frac{1}{2} \tr(\Pi_1^2 + \Pi_2^2)
    \qquad
    \Rightarrow
    \qquad
    \frac{1}{2 N^2} P \Omega_{+} P + \frac{N^2}{8} \omega_{+}^{} \Omega_{+}^{-1} \omega_{+}^{} \, ,
\end{equation}
where $P_{a} \equiv - i \partial / \partial \phi^a$ is the canonical conjugate momentum of $\phi^a$.

\paragraph{}
For the second kinetic term, we can write it as
\begin{equation}
    \tr(\Pi_1 \Pi_2) = - \frac{1}{2} \tr(\pdv[2]{}{M_1}{M_2} + \pdv[2]{}{M_2}{M_1}) \, .
\end{equation}
The resulting loop joining $\Omega_{\rm mix}$ and the loop splitting vector $\omega_{\rm mix}$ are then 
\begin{align}
    \Omega_{\rm mix} & = \Omega_{12} + \Omega_{21} \, , \\
    \omega_{\rm mix} & = 2 \, \omega_{12} \, .
\end{align}
The second kinetic term in the collective field representation then is 
\begin{equation}
    \tr(\Pi_1 \Pi_2) 
    \qquad
    \Rightarrow
    \qquad
    \frac{1}{2 N^2} P \Omega_{\mathrm{mix}} P + \frac{N^2}{8} \omega_{\mathrm{mix}} \Omega_{\mathrm{mix}}^{-1} \omega_{\mathrm{mix}} \, .
\end{equation}
Putting these pieces together, we then have the collective field (loop space) representation of the average energy operator \cref{eq:E_beta} as
\begin{equation} \label{eq:E_beta_col}
    \mathcal{E}_{\theta , \, \col}[P, \phi] = 
    \frac{1}{4 N^2} \left[
        \cosh(2 \theta(\beta)) P \Omega_{+} P + \sinh(2 \theta(\beta)) P \Omega_{\rm mix} P
    \right]
    + N^2 \mathcal{V}_{\beta , \, \col}[\phi] \, ,
\end{equation}
where we have the collective potential
\begin{equation} \label{eq:V_beta_col}
    \mathcal{V}_{\theta , \, \col}[\phi] = 
    \frac{1}{16} \left[
        \cosh(2 \theta(\beta)) 
        \big( \omega_{+}^{} \Omega_{+}^{-1} \omega_{+}^{} + 4 \phi^{3} + 4 \phi^{5} \big)
        +
        \sinh(2 \theta(\beta)) 
        \big( \omega_{\rm mix}^{} \Omega_{\rm mix}^{-1} \omega_{\rm mix}^{} - 8 \phi^{4} \big) 
    \right] \, .
\end{equation}

\paragraph{}
Similarly, the collective field representation of the average energy operator \cref{eq:E} without $G$-transformation is
\begin{equation}
    \hat{\mathcal{E}}[P, \phi] = \frac{1}{4 N^2} P \Omega_{+} P + N^2 \hat{\mathcal{V}}[\phi] \, , 
\end{equation}
with a collective potential
\begin{equation} \label{eq:V_col_TFD}
    \hat{\mathcal{V}}[\phi] = \frac{1}{16} \omega_{+}^{} \Omega_{+}^{-1} \omega_{+}^{} + \frac{1}{4} (\phi^{3} + \phi^{5}) \, .
\end{equation}
In the limit of large $N$, we would like to numerically minimize $\mathcal{V}_{\theta, \, \col}[\phi]$ \eqref{eq:V_beta_col} to solve for the loop values at finite temperature. We expect that the minimum of $\mathcal{V}_{\theta, \, \col}[\phi]$ is $1/2$, i.e. the energy at the zero temperature (ground state energy). On the other hand, the obtained loop values are those at the inverse temperature $\beta$. To compute the average energy at finite temperature, we should substitute the loop values at finite temperature, obtained from the numerical minimization, into $\hat{\mathcal{V}}[\phi]$ \eqref{eq:V_col_TFD}, i.e. the collective potential without being transformed by $G$.

\begin{figure}[thb!]
    \begin{center}
        \begin{subfigure}[c]{0.495\linewidth}
            \includegraphics[width=\textwidth]{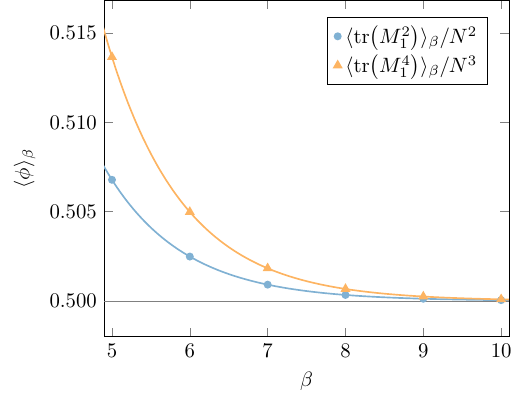}
        \end{subfigure}
        \hfill
        \begin{subfigure}[c]{0.495\linewidth}
            \includegraphics[width=\textwidth]{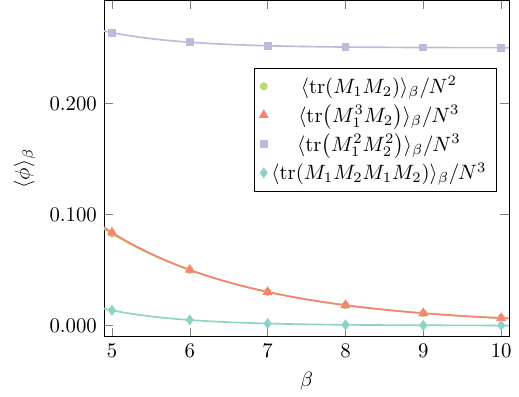}
        \end{subfigure}
        \caption{Loop values versus the inverse temperature $\beta$.}
        \label{fig:loopvalues_TFD}
    \end{center}
    \end{figure}

\paragraph{}
Let us now present some numerical results obtained by \texttt{Mathematica}. As a consistent test, we use $\ell_{\max} = 5$ so that $L_{\max} = 2 \ell_{\max} - 2 = 8$. According to \cite{Koch:2021yeb}, the sizes of $\Omega_{+}$ and $\Omega_{\mathrm{mix}}$ are both $N_{\Omega} = 23$, and the total number of loops involved is $N_{\operatorname{Loops}} = 93$. We choose $N = 10$ so that $N(N+1) = 110 > N_{\operatorname{Loops}}$. As for the inverse temperature, we fix $\beta = 5, 6, \dots , 10$. The numerical minimizations take only several minutes for these cases. The results are presented in the following figures and tables. In \Cref{fig:E_vs_beta} we plot the thermal average energy $\langle E \rangle_{\beta}$ versus $\beta$. As shown in the figure, the numerical results match perfectly with the analytical calculations. We also tested the invariants $4(\phi^3_{\beta} \phi^5_{\beta} - (\phi^4_{\beta})^2) = 1$ based on the numerical results, and indeed they all give $1$. In \Cref{fig:loopvalues_TFD} we plot the loops in $V_{2}$ and $V_{4}$ versus $\beta$, exhibiting perfect agreement with analytical results. 

\begin{figure}[thb!]
    \begin{center}
        \includegraphics[width=0.5\textwidth]{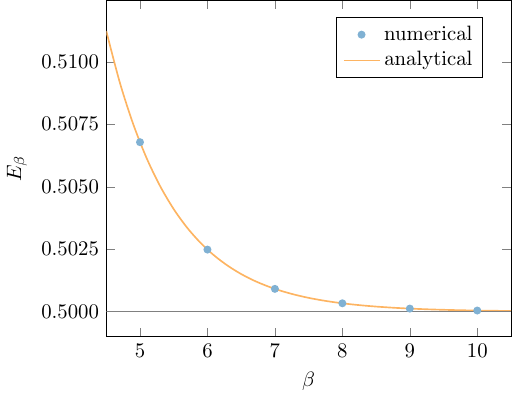}
        \caption{Thermal energy versus the inverse temperature $\beta$.}
        \label{fig:E_vs_beta}
    \end{center}
    \end{figure}

\subsection{Symmetry and zero modes}

We then consider the $1/N$-expansion of the $\hat{G}$ operator. For free theory, the $\hat{G}$ operator in loop space reads
\begin{equation} \label{eq:G_2_loop}
    \hat{G} = \sum_{C} \Omega_{+}(M_1 M_2, C) P(C) \, .
\end{equation}
The $1/N$-expansion 
\begin{equation}
    \phi(C) = \phi_{\theta}(C) + \frac{1}{N} \eta(C) \, , 
    \qquad
    P(C) = N p(C)
\end{equation}
gives
\begin{equation}
    \hat{G} = N \sum_{C} \Omega_{+, \theta}(M_1 M_2, C) p(C) + \sum_{C} \Omega_{+, \theta}(M_1 M_2, C)_{\phi \rightarrow \eta} \, p(C) \, .
\end{equation}
Similar to the O($N$) vector TFD \cite{Jevicki:2023yuh}, the leading term is also of order $N$. Here $\Omega_{+, \theta}$ denotes the loop joining $\Omega_{+}$ at the thermal background, and is identical to the zero mode 
\begin{equation}
    v_{\theta}(C) = \Omega_{+, \theta}(M_1 M_2, C) = \pdv{\phi_{\theta}(C)}{\theta} \, .
\end{equation}
Writing $\hat{G} = N \hat{G}^{(1)} + \hat{G}^{(2)}$, we see the order $N$ term can be written as
\begin{equation}
    \hat{G}^{(1)} = \sum_{C} v_{\theta}(C) p(C) \, .
\end{equation}
We note that $v_{\theta}$ is a zero mode of $\hat{V}_{\theta}^{(2)}$:
\begin{equation}
    \sum_{C^\prime} \hat{V}_{\theta}^{(2)}(C, C^\prime) v_{\theta}(C^\prime) = 0 \, ,
\end{equation}
where $\hat{V}_{\theta}^{(2)}$ is the second derivative matrix of the collective potential $\hat{V}_-$ at the thermal background $\phi_{\theta}$.

\paragraph{}
Let us then consider the $1/N$-expansion of the Hamiltonian $\hat{H}_{+} \equiv 2 \hat{\mathcal{E}}$. We have the $1/N$-expansion
\begin{equation}
    \hat{H}_{+} = N^2 \frac{\cosh(2 \theta_{\beta})}{2} + N \hat{H}_{+}^{(1)} + \hat{H}_{+}^{(2)} + \frac{1}{N} \hat{H}_{+}^{(3)} + \dots \, .
\end{equation}
The order $N^2$ term is the leading thermal average energy. The order $N$ term is given by
\begin{equation}
    \hat{H}_{+}^{(1)} = \sum_{C} u_{\theta}(C) \eta(C) \, ,
    \quad
    u_{\theta}(C) \equiv \left. \pdv{\hat{V}_+}{\phi(C)} \right|_{\phi_{\theta}} \, .
\end{equation}
%
%
We note that $u_{\theta}$ is a zero mode of $\Omega_{-, \theta}$:
\begin{equation}
    \sum_{C^\prime} \Omega_{-, \theta}(C, C^\prime) u_{\theta}(C^\prime) = 0 \, .
\end{equation}

\paragraph{}
The commutator $[\hat{H}_+, \hat{G}]$ gives
\begin{equation}
    [\hat{H}_+, \hat{G}] = N^2 [\hat{H}_+^{(1)}, \hat{G}^{(1)}] + \mathcal{O}(N) \, ,
\end{equation}
where the leading order term gives
\begin{equation}
    [\hat{H}_+^{(1)}, \hat{G}^{(1)}] = \ii \sum_{C} u_{\theta}^{\rm T}(C) v_{\theta}(C) = 2 \ii \sinh(2 \theta_{\beta}) \, .
\end{equation}
This reveals that $\hat{G}^{(1)}$ can be regarded as the canonical conjugate of $\hat{H}_+^{(1)}$. One can again perform similar analysis using the collective coordinate method \cite{Jevicki:2023yuh}.

\section{Interacting theory}
\label{sec:Interacting_theory}

We now consider the more general interacting theory case. 
To start we present a discussion of extending the Bogoliubov symmetry from the free case. Based on our previous solution of  vector type theories we expect that such extension is operational on-shell. This interacting ``dynamical symmetry" is furthermore state dependent, much like the Kubo–Martin–Schwinger (KMS) conditions that apply to thermal expectation values.

\subsection{Dynamical symmetry}

In free theory case we have the $G$ operator \eqref{eq:G} which commutes with the free thermofield Hamiltonian. For interacting theory case, we can construct it from the symmetry condition, representing a dynamical symmetry. The procedure is similar to the O($N$) vector model case \cite{Jevicki:2021ddf}, which we will illustrate in detail for matrix models. Let us then consider generic interacting theories ($M_1 \equiv M, \, M_2 \equiv \tilde{M}$)
\begin{equation}
    H_1 = \frac{1}{2} \tr(\Pi_1^2) + \frac{\omega_1^2}{2} \tr(M_1^2) + \frac{g_n}{N^{\frac{n}{2} - 1}} \tr(M_1^n) \, , 
    \qquad n \geq 3 \, .
\end{equation}
The thermofield Hamiltonian is again $H_{-} = H_1 - H_2$. We put back the frequencies $\omega_1$ and $\omega_2$ since they could play a role for the regularization of the $G_{f}$ operator \cite{Jevicki:2021ddf}. We would like to solve for $G_f$ from the symmetry condition:
\begin{equation} \label{eq:H_G_constraint}
    [H_{-}, G_{f}] = 0 \, .
\end{equation}
We can first separate the free parts and the interacting parts of the operators:
\begin{equation}
    H_{-} = H^{(2)}_{-} + \frac{g_n}{N^{\frac{n}{2}-1}} H^{(n)}_{-} \, , \qquad 
    H^{(n)}_{-} = \tr(M_1^n) - \tr(M_2^n) \, .
\end{equation}
Here $H^{(2)}_{-}$ represents the free theory thermofield Hamiltonian \cref{eq:H_hat_2}. Similarly for $G_{f}$ we have
\begin{equation}
    G_{f} = G_{f}^{(2)} + \frac{g_n}{N^{\frac{n}{2}-1}} G_{f}^{(n)} \, ,
\end{equation}
with $G_{f}^{(2)}$ in \eqref{eq:G} ($f = \theta(\beta)$) and $G_{f}^{(n)}$ to be solved. Substituting these relations into the constraint \cref{eq:H_G_constraint}, we have in the weak coupling limit
\begin{equation} \label{eq:G_n_weak_coupling_constraint}
    [H^{(2)}_{-}, G_{f}^{(n)}] = [G_{f}^{(2)}, H^{(n)}_{-}] \, .
\end{equation}
On the other hand, in the strong coupling limit, we have
\begin{equation} \label{eq:G_n_strong_coupling_constraint}
    [H^{(n)}_{-}, G_f^{(n)}] = 0 \, .
\end{equation}

\paragraph{}
In the weak coupling limit, \cref{eq:G_n_weak_coupling_constraint} is equivalent to
\begin{equation} \label{eq:G_n_weak_coupling_constraint_expansion}
    \frac{1}{2} [\tr(\Pi_1^2 + \omega_1^2 M_1^2) - \tr(\Pi_2^2 + \omega_2^2 M_2^2), G_{f}^{(n)}]
    = n \ii f \big(\tr(M_1 M_2^{n-1}) - \tr(M_1^{n-1} M_2)\big) \, .
\end{equation}
One can use \texttt{Mathematica} to solve for this equation, and below we present a brief algorithm. At first we observe that $G_{f}^{(n)}$ must be summations of single-trace operators with $n$ matrices, which could involve $M_1, M_2, \Pi_1$, and $\Pi_2$. Thus, we need to generate all such loop variables, given by the set
\begin{equation}
    U_{G} = \frac{U}{\mathbb{Z}_{n}} \, ,
    \qquad
    U = \{ \tr(\mathcal{O}_1 \cdots \mathcal{O}_n)\mid \mathcal{O}_i \in \{ M_1, M_2, \Pi_1, \Pi_2 \} \, , i = 1, \dots, n \} \, ,
\end{equation}
where $\mathbb{Z}_{n}$ is the cyclic group. At the quantum level, there is also an ordering problem, e.g., $\operatorname{tr}(M_1 \Pi_1) \neq \tr(\Pi_1 M_1)$. To resolve this issue we can perform a symmetrization in the end. At this stage we can simply ignore it. The $G_{f}^{(n)}$ operator will be a linear combination of the loops in~$U_{G}$
\begin{equation}
    G_{f}^{(n)} = f \sum_{r} c_{r} \, \varphi^{r} \, , \qquad \varphi^{r} \in U_{G} \, .
\end{equation}
Our goal is to solve for the linear coefficients $c_{r}$ from the constraint equation. Since $H^{(2)}_{-}$ is a quadratic form of the matrices, we have
\begin{equation}
    [H^{(2)}_{-}, \varphi^{n}] = \sum_{s} h^{r}{}_{s} \, \varphi^{s} \in \vspan(U_{G}).
\end{equation}
Substituting this linear expansion into \cref{eq:G_n_weak_coupling_constraint_expansion} we then obtain a set of linear equations for~$c_{r}$:
\begin{equation} \label{eq:c_equations}
    \sum_{r, s} h^{r}{}_{s} c_{r} \varphi^{s} = n \ii \big(\tr(M_1 M_2^{n-1}) - \tr(M_1^{n-1} M_2)\big) \, .
\end{equation}
We should then compute the commutators of $H^{(2)}_{-}$ with all loops in $U_{G}$ to determine the matrix~$h^{r}{}_{s}$. This can be simply implemented in computer programs as follows. As in the loop word representation we respectively label $M_{1}$ and $M_{2}$ with numbers 1 and 2, we can also label $\Pi_1$ and $\Pi_2$ with numbers 3 and 4, respectively. The loop $\tr(M_1 M_2 M_1 \Pi_1)$, taking $n=4$ for example, is then represented as an array (list) $\{1, 2, 1, 3\}$. To compute the commutators of $H^{(2)}_{-}$ with them, we simply replace the numbers with those that represent their conjugates. For example, if we would like to implement the commutator of $\tr(M_1^2)$, which is represented by $\{1, 1\}$, with the loop $\tr(M_1 M_2 M_1 \Pi_1)$ mentioned above, we simply replace the number $3$ in that loop with $1$, such that we have $\{1, 2, 1, 1\}$. Since there are two $M_1$ in $\tr(M_1^2)$, the final result is represented by two copies of $\{1, 2, 1, 1\}$. It is not necessary to incorporate the factor $\ii$ since this imaginary unit will be canceled out on the right hand side of \cref{eq:G_n_weak_coupling_constraint_expansion}. We also have to multiply with the factor $-1$ if we compute the commutators with $\tr(\Pi_1^2)$ and $\tr(\Pi_2^2)$. 

\paragraph{}
Once the matrix $h^{r}{}_{s}$ is determined, we can substitute it into the constraint \cref{eq:c_equations} to solve for the coefficients~$c_r$. We expect that the solution is unique for a specific $n$. This then completes the form of $G_{f}$ in the weak coupling limit. If we have a unique solution, we can substitute it into the constraint for the strong coupling limit \cref{eq:G_n_strong_coupling_constraint}. If the commutator indeed vanishes, then we can safely claim that the solution actually applies for all interaction couplings. This consistency check then lifts the weak coupling result to a non-perturbative result.

\paragraph{}
We now present the simplest interacting theory: the cubic interaction with $n=3$. In this case we find that $U_{G}$ contains 24 loops. Using \texttt{Mathematica} we find that in the weak coupling limit, there is indeed a unique solution for $G_{f}^{(3)}$. The result is
\begin{align}
    G_{f}^{(3)} =  f \Big( &
        \tr(M_1^2 \Pi_2) + \tr(M_2^2 \Pi_1) + 2 \tr(\Pi_1^2 \Pi_2) + 2 \tr(\Pi_2^2 \Pi_1) \nonumber \\ &
        - \tr(M_1 M_2 \Pi_1) - \tr(M_2 M_1 \Pi_1) - \tr(M_2 M_1 \Pi_2) - \tr(M_1 M_2 \Pi_2) \Big) \, \nonumber \\
        + & (\text{symmetrization}) \, .
\end{align}
We see that $G_{f}^{(3)}$ also has the $\mathbb{Z}_2$ symmetry as in $G_{f}^{(2)}$.

\paragraph{}
Substituting the solution into the strong coupling limit constraint, we obtain
\begin{align}
    [H^{(3)}_{-}, G_{f}^{(3)}] = 6 \ii f \Big( &
        - \tr(M_1^3 M_2) + \tr(M_1^2 \Pi_1 \Pi_2) + \tr(M_1^2 \Pi_2 \Pi_1) + \tr(M_2^2 \Pi_1^2) \nonumber \\ &
        + \tr(M_2^3 M_1) - \tr(M_2^2 \Pi_2 \Pi_1) - \tr(M_2^2 \Pi_1 \Pi_2) + \tr(M_1^2 \Pi_2^2) \Big) \nonumber \\
        + & (\text{symmetrization}) \, .
\end{align}
Although the commutator does not vanish automatically, due to the $\mathbb{Z}_2$ symmetry that exchanges the matrix indices, those loops that are related by this symmetry should have the same expectation value. For example, $\langle\tr(M_1^3 M_2)\rangle = \langle\tr(M_2^3 M_1)\rangle$. We see that terms on the right hand side cancel out when taking expectation values. In this sense we can claim that the solution we obtain in the weak coupling limit also solves the constraint in the strong coupling limit.

\subsection{Numerical optimization scheme}

To be concrete, we consider the interacting theory with a quartic single-trace operator. The Hamiltonian governing the time evolution is \cref{eq:H_hat}, $H_- = H_1 - H_2$, representing two copies of MQM with a quartic single-trace interaction:  
\begin{equation}
    H = \tr(\frac{1}{2} \Pi^2 + \frac{1}{2} M^2 + \frac{g}{N} M^4) \, .
\end{equation}
We then have the thermofield collective Hamiltonian
\begin{equation*}
    \hat{H}_-[P, \phi] = \frac{1}{2 N^2} P(C_1) \Omega_-(C_1, C_2) P(C_2) + N^2 \hat{V}_-[\phi] \, ,
\end{equation*}
where the thermofield collective potential is
\begin{equation}
\begin{split}
    \hat{V}_-[\phi] &= \frac{1}{8} \omega_-(C_1) \Omega_-^{-1}(C_1, C_2) \omega_-(C_2) + \frac{1}{2} \phi(M_1^2) - \frac{1}{2} \phi(M_2^2) + g \, \phi(M_1^4) - g \, \phi(M_2^4) \\
    &= \frac{1}{8}\omega_-^{\rm{T}}\Omega_-^{-1}\omega_- + \frac{1}{2}(\phi_3-\phi_5) + g(\phi_{10} - \phi_{15}) \, .
\end{split}
\end{equation}
We would like to perform a ``master field optimization" \cite{Koch:2021yeb} to solve for the large $N$ background.  In this case we do not have the exact formula for $G$, and consequently we do not know the $G$-transformed $H_{+}$ as in the free case. However, based on our analytical study of vector models, the problem will be addressed in the real time Hilbert space with the Hamiltonian $H_{-}$. The possibility that the thermal solution can be obtained solely based on $H_{-}$ comes from identification of a nonlinear symmetry governing the thermal trajectory. To avoid the unboundedness of the Hamiltonian $H_{-}$ we will introduce another quantity that is amenable for minimization. We first note that, due to the anti-symmetry that $\hat{V}_- \rightarrow - \hat{V}_-$ under the $\mathbb{Z}_2$ transformation $M_1 \leftrightarrow M_2$, if one performs a ``Keldysh rotation''
\begin{equation}
    X \equiv M_1 + M_2 \, , \quad 
    Y \equiv M_1 - M_2 \, ,
\end{equation}
then all loops consisting of odd number of $Y$s should vanish:
\begin{equation}
\label{eq:odd_cstr}
    \phi(C) = 0 \quad \text{if } \# \, Y {\text{ in }} C {\text{ is odd}} \, .
\end{equation}
On the other hand, though the thermofield collective potential $\hat{V}_-$ is unbounded from below, its derivative with respect to the master fields must vanish at the large $N$ background. Besides, based on our previous knowledge of the TFD states, there should exist infinitely many solutions, each corresponding to a thermal background with different temperatures. To obtain a unique solution we should fix one collective field. It is simplest to fix $\phi(M_1 M_2)$. Collecting these properties, we can write down the following function $A$ for the master field optimization:
\begin{equation} \label{eq:fM}
    A = \sum_{i=1}^{2} \tr(\pdv{\hat{V}_-}{M_i} \pdv{\hat{V}_-}{M_i})
    + \big(\phi(M_1 M_2) - y_0\big)^2
    + \sum_{\substack{C \\ \# \, Y {\text{odd}}}} \Big( \phi(C) \Big)^2 \, .
\end{equation}
The first term demands that the gradient of $\hat{V}_-$ vanishes at the large $N$ background. The second term fixes the values of $\phi(M_1 M_2)$ to $y_0$ to produce a unique solution. The last term ensures that the loops with odd number of $Y$ vanish. We can then minimize the objective function $A$ using the numerical framework in \cite{Koch:2021yeb} to obtain the large $N$ thermal backgrounds. We note that the first term requires a gradient calculation, which is much more computational costly than minimizing $\hat{V}_{+}$ in the work \cite{Koch:2021yeb}. Nevertheless we can still perform the numerical optimization as it will be demonstrated.

\subsection{Numerical results}

In the previous sections, we have established the framework for thermofield minimization, at heart of which is a conjectured dynamical symmetry of thermal theories. In this subsection, we can explicitly observe this symmetry from the numerical framework in the example of interacting matrix quantum mechanics. Subsequently, the thermal loop values and the thermal energy can be extracted by master field optimization. Based on the obtained results, normal modes, $O(1)$ fluctuation and correlators follow. We first provide a refined numerical scheme that is more suitable for numerical optimization, and then present the numerical results.

\paragraph{}
At the large $N$ background $\phi_{\beta}$, the thermofield collective potential vanishes and the associated equations of motion are obeyed:
\begin{equation}
\label{eq:eoms}
    \hat{V}_-(\phi_{\beta}) = 0 \, , \quad 
    \left.
    \Omega_{+}(C, C') \pdv{\hat{V}_-}{\phi(C')}
    \right|_{\phi = \phi_{\beta}} = 0 \, ,
\end{equation}
where the derivative of $\hat{V}_{-}$ with respect to loops is
\begin{equation}
    \pdv{\hat{V}_-}{\phi_i} = \frac{1}{4}\pdv{\omega_-^{\rm{T}}}{\phi_i} Q_- - \frac{1}{8} Q_-^{\rm{T}} \pdv{\Omega_-}{\phi_i} Q_- + \frac{1}{2}\left(\delta_{i,3} - \delta_{i,5}\right) + g\left(\delta_{i,10} - \delta_{i,15}\right) \, , \quad Q_- \equiv \Omega_-^{-1}\omega_- \, .
\end{equation}
The $\mathbb{Z}_2$ symmetry that exchanges $M_1 \leftrightarrow M_2$ imposes the following constraint rules:
\begin{enumerate}[label=(\roman*)]
    \item Two loops related by reflection of words are equal-valued.
    \item Two loops related by exchanging $M_1$ and $M_2$ are equal-valued.
\end{enumerate}
These two constraints are equivalent to \eqref{eq:odd_cstr}, and imply the first equation of \eqref{eq:eoms}. In practice, it is more convenient to implement constraints in this way instead of performing transformation to symmetric-antisymmetric basis.  Due to the presence of the zero mode, the number of variables is greater than the number of equations of motion by one, so we also need to fix one thermal loop value to a target value. To be specific, we will fix 
\begin{equation}
y \equiv \phi_4 \equiv \phi(M_1 M_2) = y_0 > 0     
\end{equation}
to represent the temperature that the target value corresponds to.

\paragraph{}
Naively one might try to solve the equations of motion with constraints by minimizing the loss function
\begin{equation}
\label{eq:loss_1}
    A = \left(\pdv{\hat{V}_-}{\phi}\right)^{\rm{T}}\Omega_{+}\pdv{\hat{V}_-}{\phi} + (y - y_0)^2 + \sum_{\substack{C~ \text{odd}}}\phi^2(C) \, .
\end{equation}
The first term in \eqref{eq:loss_1} is a rewrite of \eqref{eq:fM}. Note that it is crucial to insert a positive semi-definite matrix $\Omega_+$ to lift the Hankel constraints. Instead of changing to the ``Keldysh basis", we find it more convenient to categorize thermal loops based on the constraint rules above before optimization, and the constraints in \eqref{eq:loss_1} are computed as the variances of observed degenerate thermal loop values during optimization. However, due to the presence of zero mode in $\Omega_-$, there are multiple solutions for $Q_-$ and it cannot be determined uniquely.

\paragraph{}
A resolution to circumvent this issue is to solve the \textit{dual} equations \eqref{eq:dual}. We only need to replace the first term of the loss function \eqref{eq:loss_1}:
\begin{equation}
\label{eq:loss_2}
    A = \norm{\Omega_{-}\pdv{\hat{V}_+}{\phi}}^2 + (y - y_0)^2 + \sum_{\substack{C~ \text{odd}}}\phi^2(C) \, .
\end{equation}
This objective function is more desirable because $\Omega_+$ does not possess any genuine zero mode and one can compute its inverse numerically without difficulty. The equations of motion becomes
\begin{equation}
    \pdv{\hat{V}_+}{\phi_i} = \frac{1}{4}\pdv{\omega_+^{\rm{T}}}{\phi_i} Q_+ - \frac{1}{8}Q_+^{\rm{T}}\pdv{\Omega_+}{\phi_i}Q_+ + \frac{1}{2}\left(\delta_{i,3} + \delta_{i,5}\right) + g\left(\delta_{i,10} + \delta_{i,15}\right) \, , \quad Q_+ = \Omega_+^{-1}\omega_+ \, .
\end{equation}

\paragraph{}
We initialize the master fields by the zero temperature solution. Apparently the equations of motion and the constraints terms are very small (they are not exact zero due to some numerical errors). Therefore the only contribution to the loss function $A$ results from the second term. During optimization, we observe that the first and third terms fluctuate around the initial small values, while the second term keeps decreasing, which indicates that the code not only searches for an interacting solution at the temperature corresponding to $y_0$, but also travels along the dynamical symmetry trajectory. In other words, every step gives a validated solution at the temperature corresponding to the associated $y$ value.

\begin{figure}[htb!]
    \centering
    \begin{subfigure}{0.3\textwidth}
        \centering
        \includegraphics[width=\linewidth]{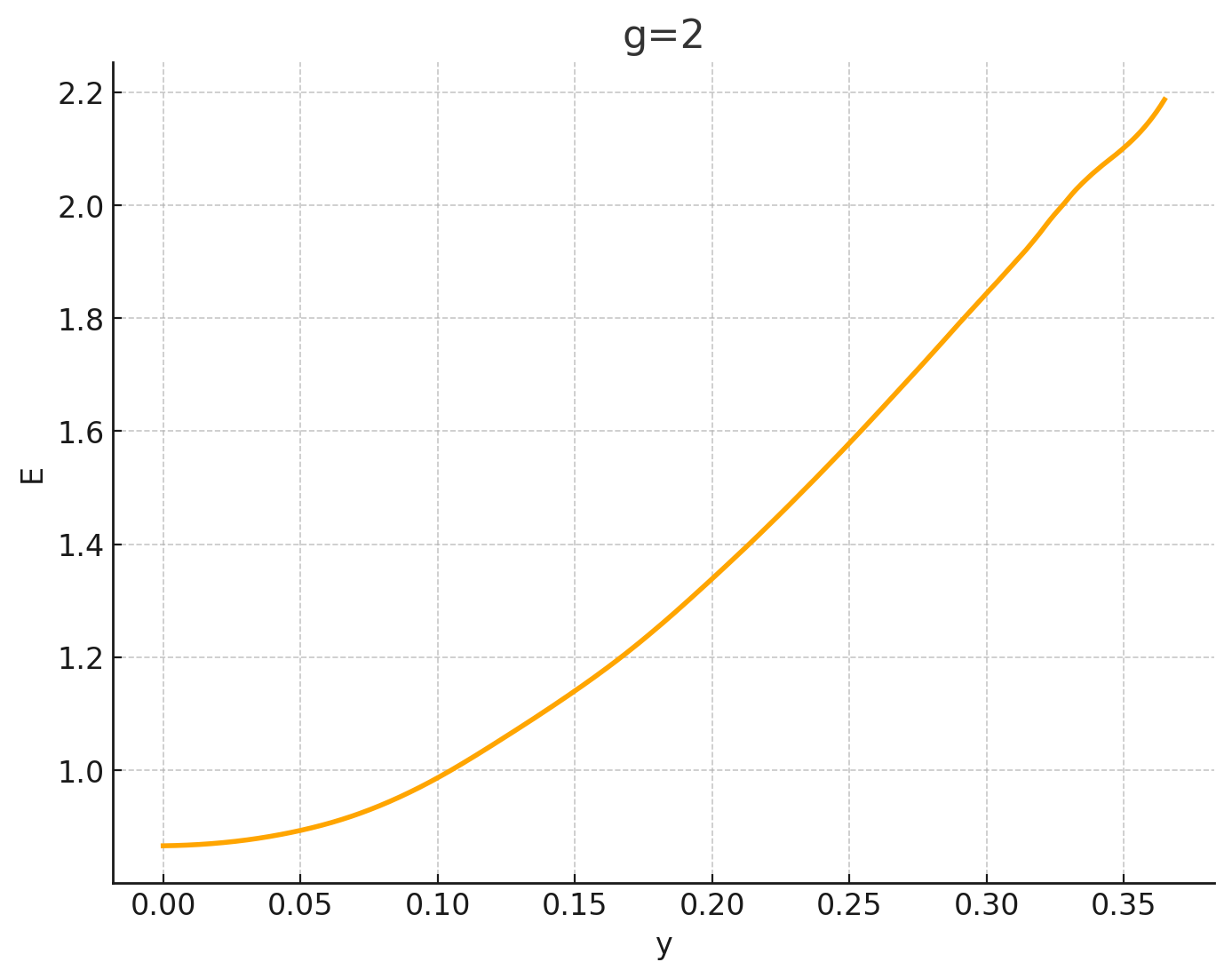}
        \caption{$E$ vs $y$, $g=2$.}
        \label{fig:E_phi4_g2}
    \end{subfigure}%
    \hfill
    \begin{subfigure}{0.3\textwidth}
        \centering
        \includegraphics[width=\linewidth]{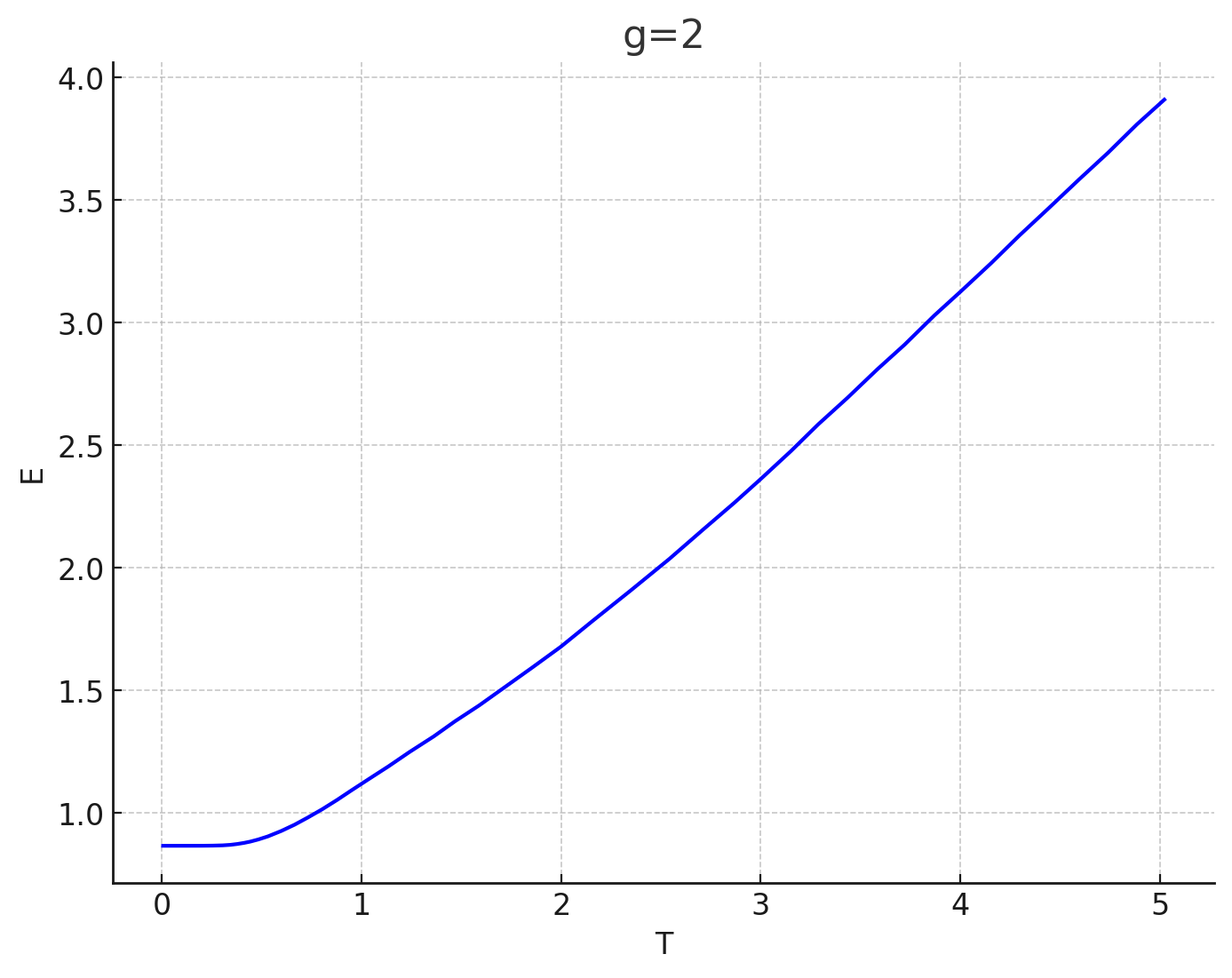}
        \caption{$E$ vs $T$, $g=2$.}
        \label{fig:E_T_g2}
    \end{subfigure}%
    \hfill
    \begin{subfigure}{0.3\textwidth}
        \centering
        \includegraphics[width=\linewidth]{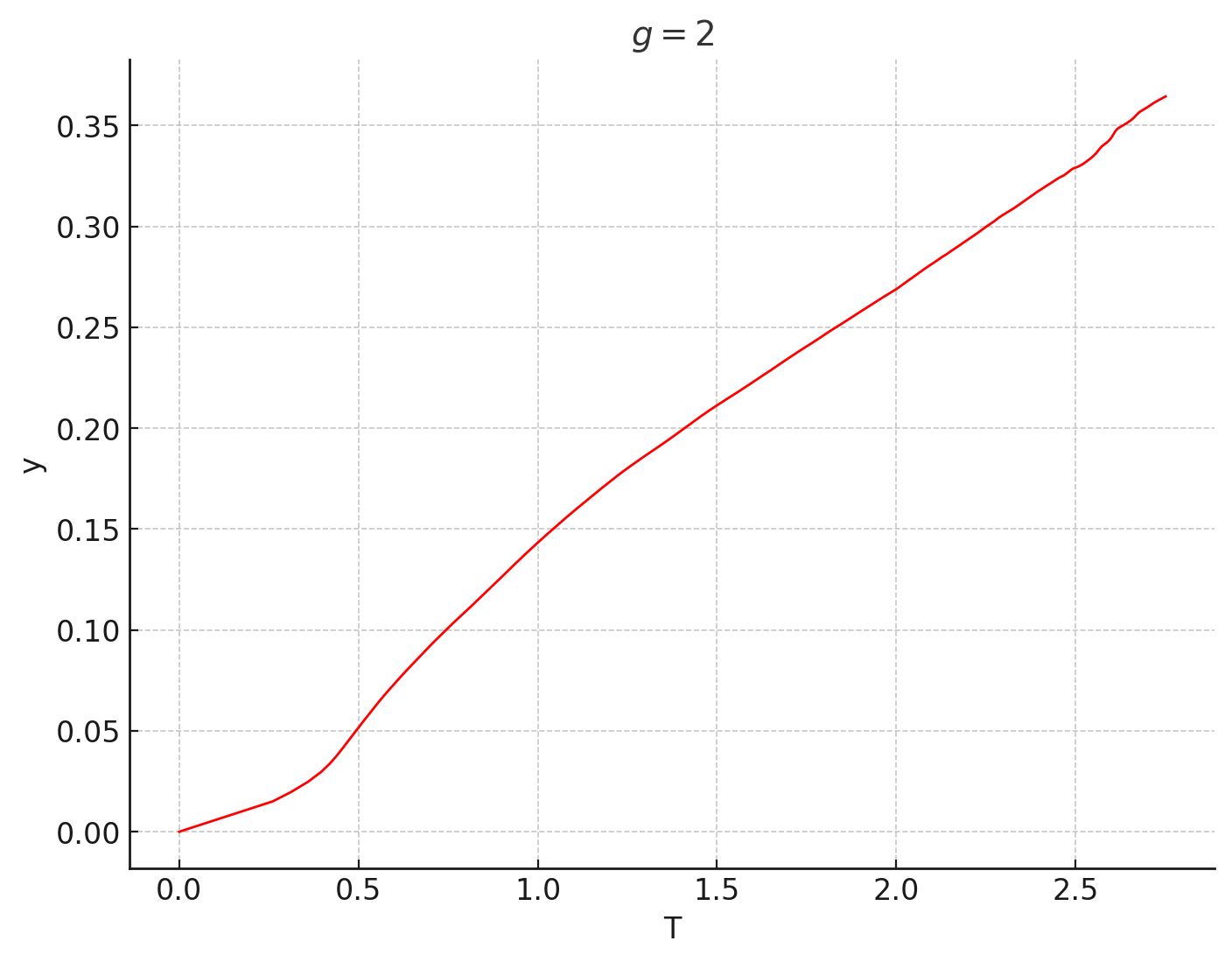}
        \caption{$y$ vs $T$, $g=2$.}
        \label{fig:phi4_T_g2}
    \end{subfigure}
    \caption{Comparison of $E$ vs $y \equiv \phi(M_1 M_2)$, $E$ vs $T$, and $y$ vs $T$ for $g=2$.}
    \label{fig:g2}
\end{figure}

\begin{figure}[htb!]
    \centering
    \begin{subfigure}{0.3\textwidth}
        \centering
        \includegraphics[width=\linewidth]{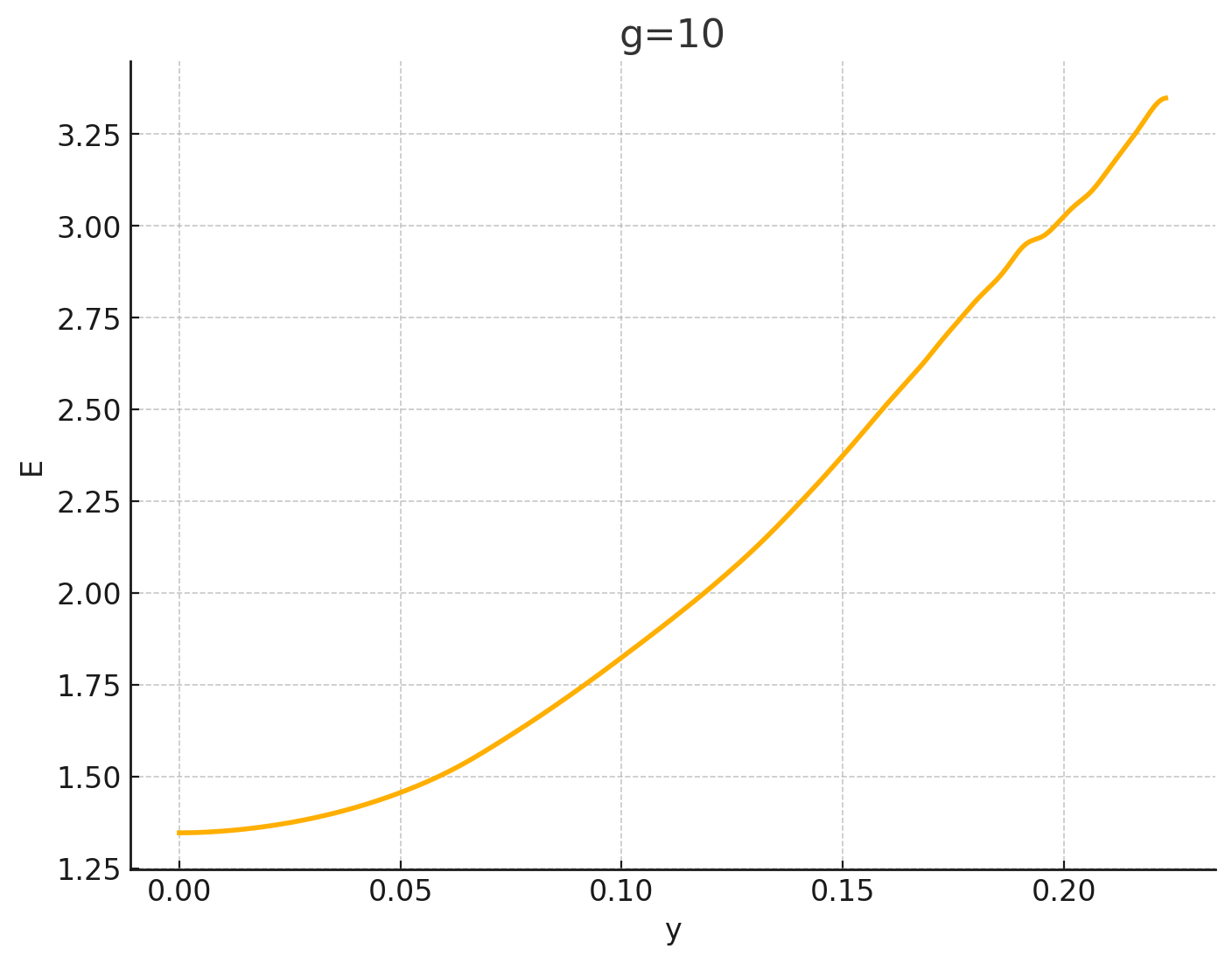}
        \caption{$E$ vs $y$, $g=10$.}
        \label{fig:E_phi4_g10}
    \end{subfigure}%
    \hfill
    \begin{subfigure}{0.3\textwidth}
        \centering
        \includegraphics[width=\linewidth]{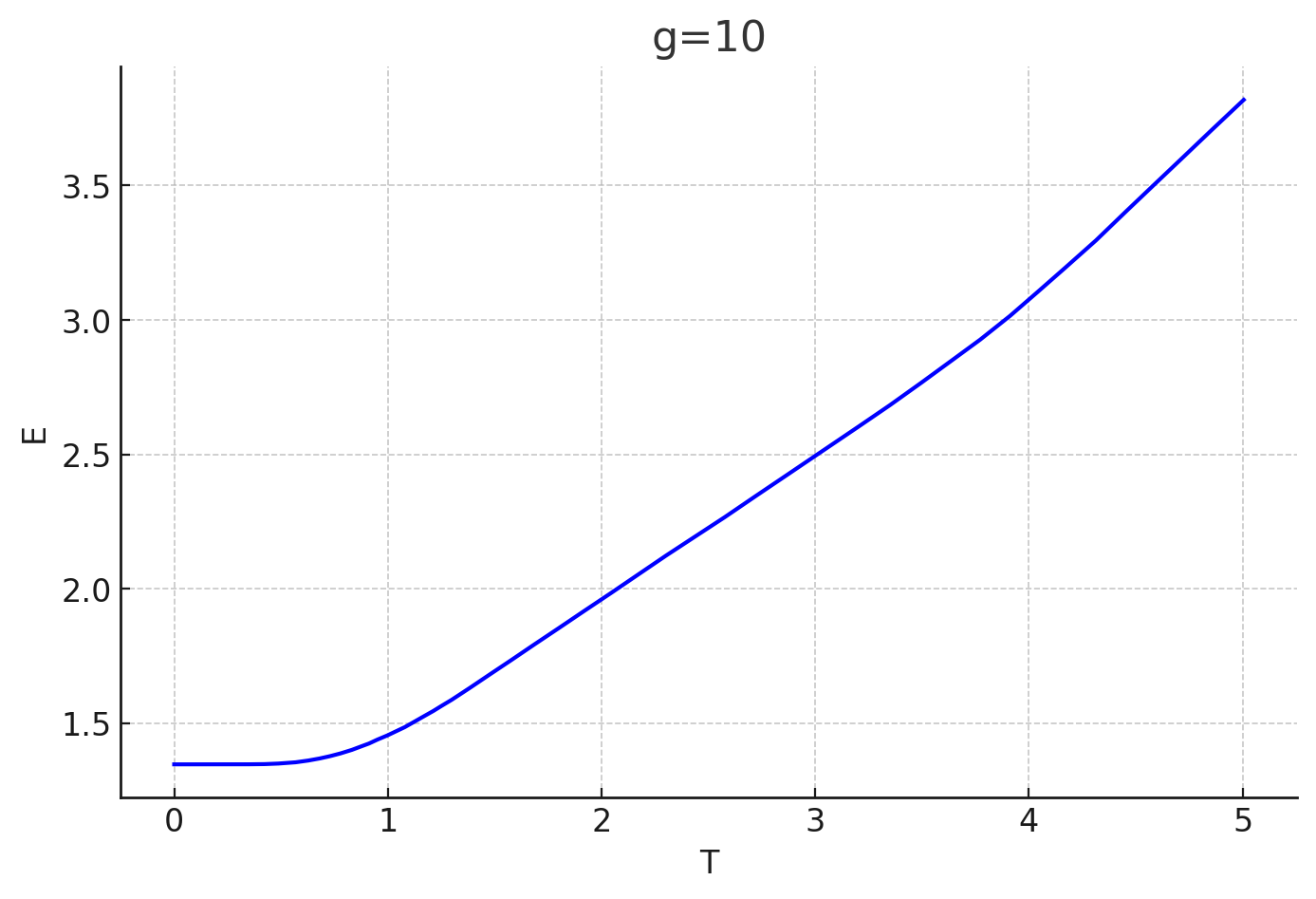}
        \caption{$E$ vs $T$, $g=10$.}
        \label{fig:E_T_g10}
    \end{subfigure}%
    \hfill
    \begin{subfigure}{0.3\textwidth}
        \centering
        \includegraphics[width=\linewidth]{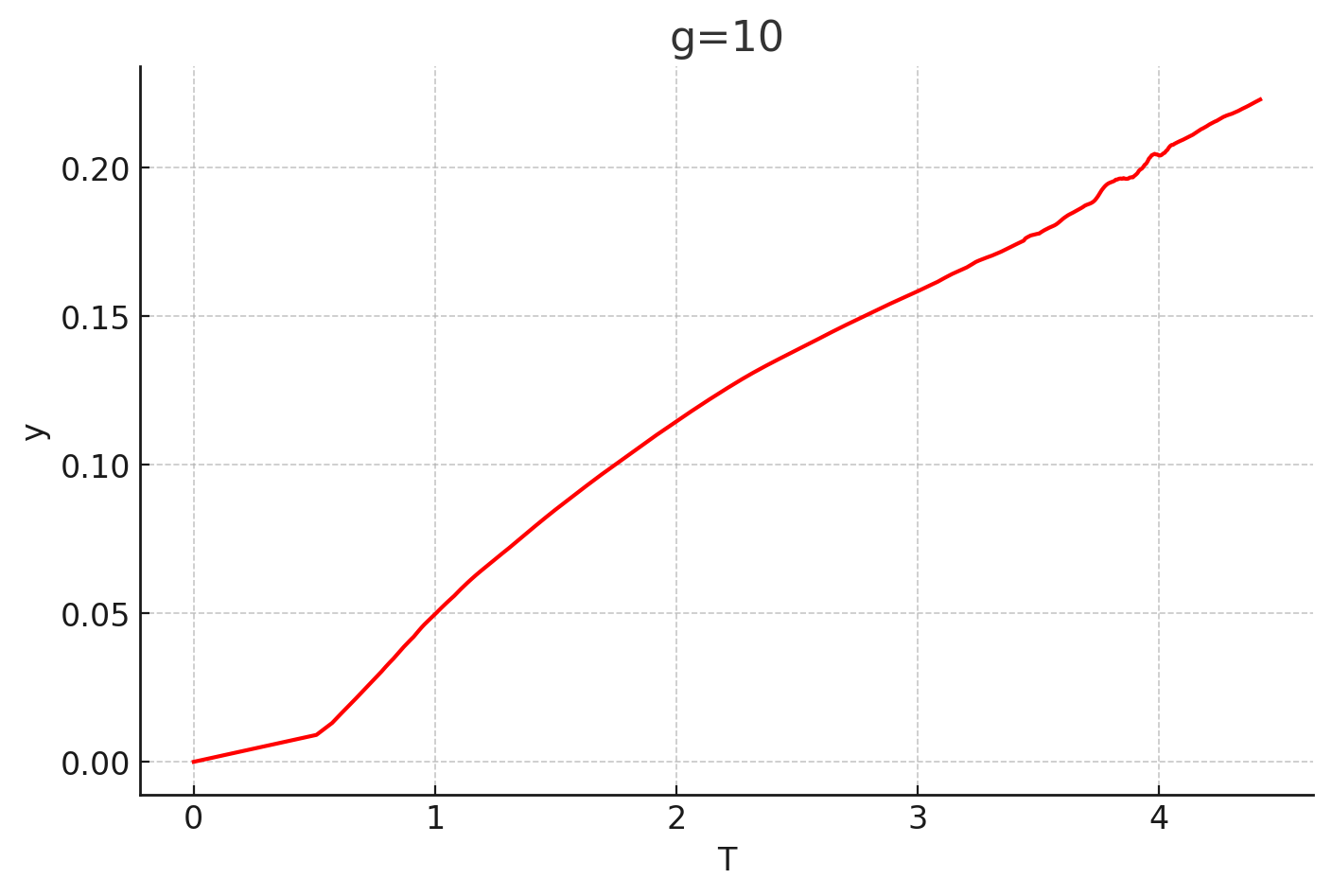}
        \caption{$y$ vs $T$, $g=10$.}
        \label{fig:phi4_T_g10}
    \end{subfigure}
    \caption{Comparison of $E$ vs $y \equiv \phi(M_1 M_2)$, $E$ vs $T$, and $y$ vs $T$ for $g=10$.}
    \label{fig:g10}
\end{figure}

\begin{figure}[htb!]
    \centering
    \begin{subfigure}{0.3\textwidth}
        \centering
        \includegraphics[width=\linewidth]{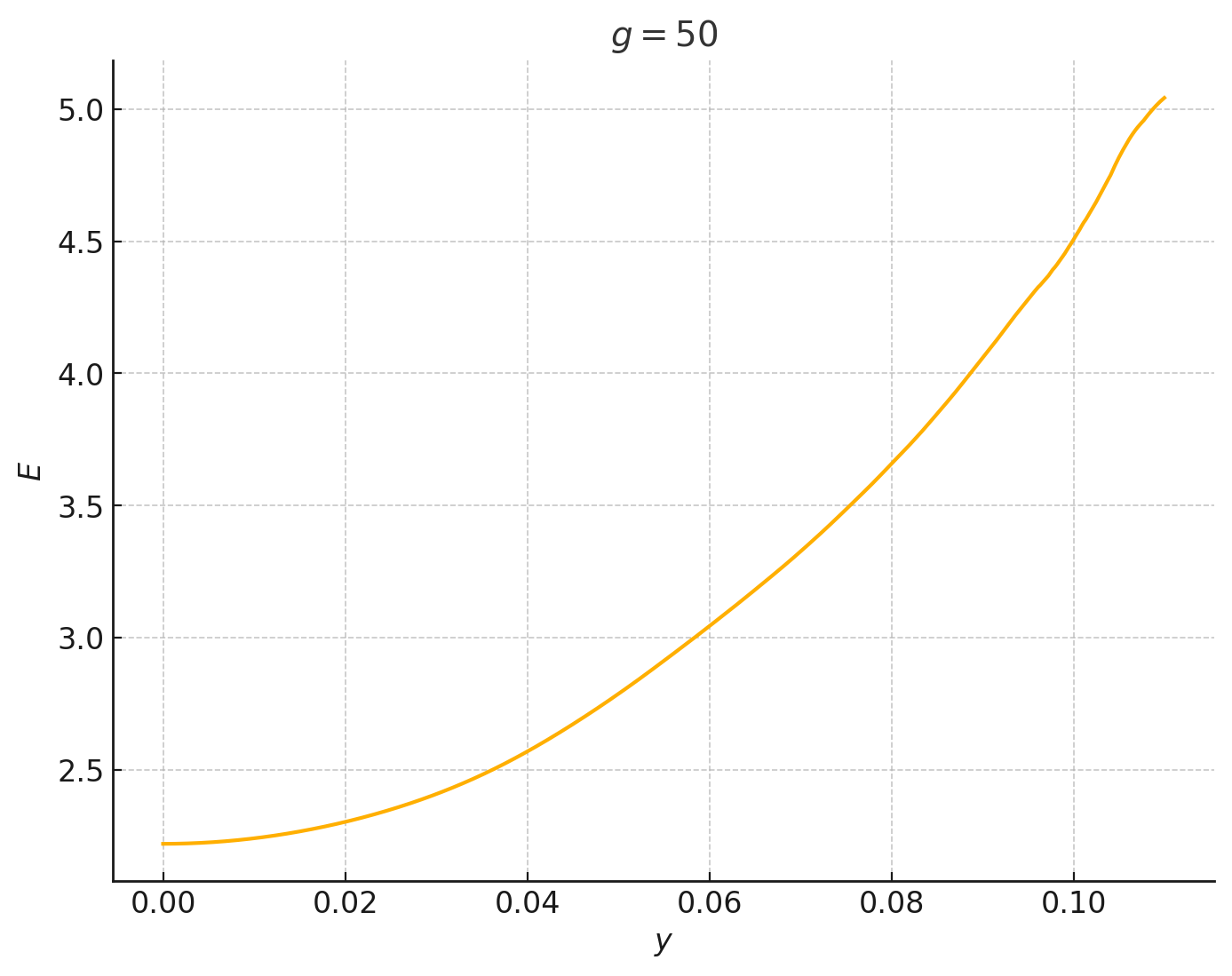}
        \caption{$E$ vs $y$, $g=50$.}
        \label{fig:E_phi4_g50}
    \end{subfigure}%
    \hfill
    \begin{subfigure}{0.3\textwidth}
        \centering
        \includegraphics[width=\linewidth]{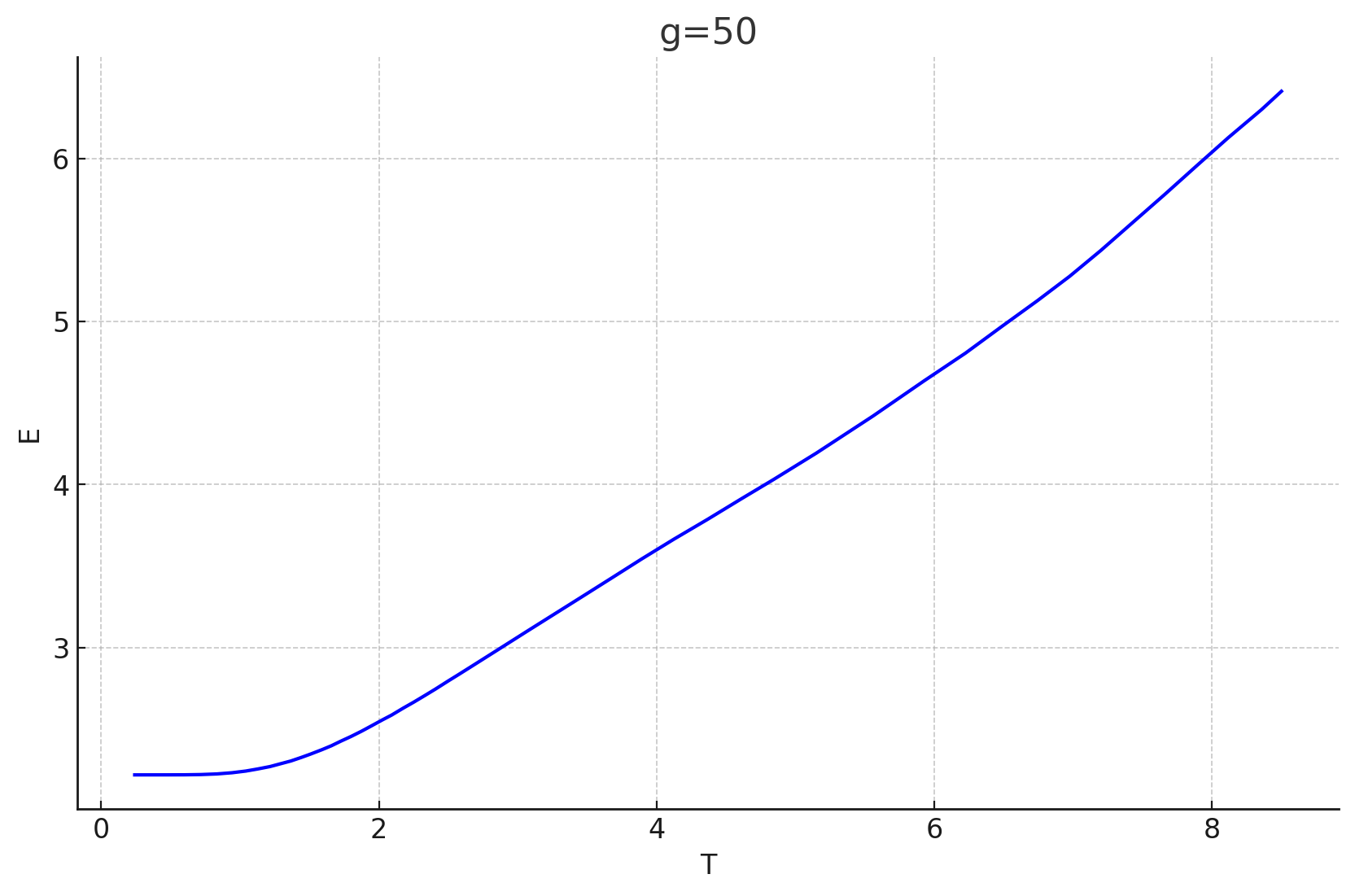}
        \caption{$E$ vs $T$, $g=50$.}
        \label{fig:E_T_g50}
    \end{subfigure}%
    \hfill
    \begin{subfigure}{0.3\textwidth}
        \centering
        \includegraphics[width=\linewidth]{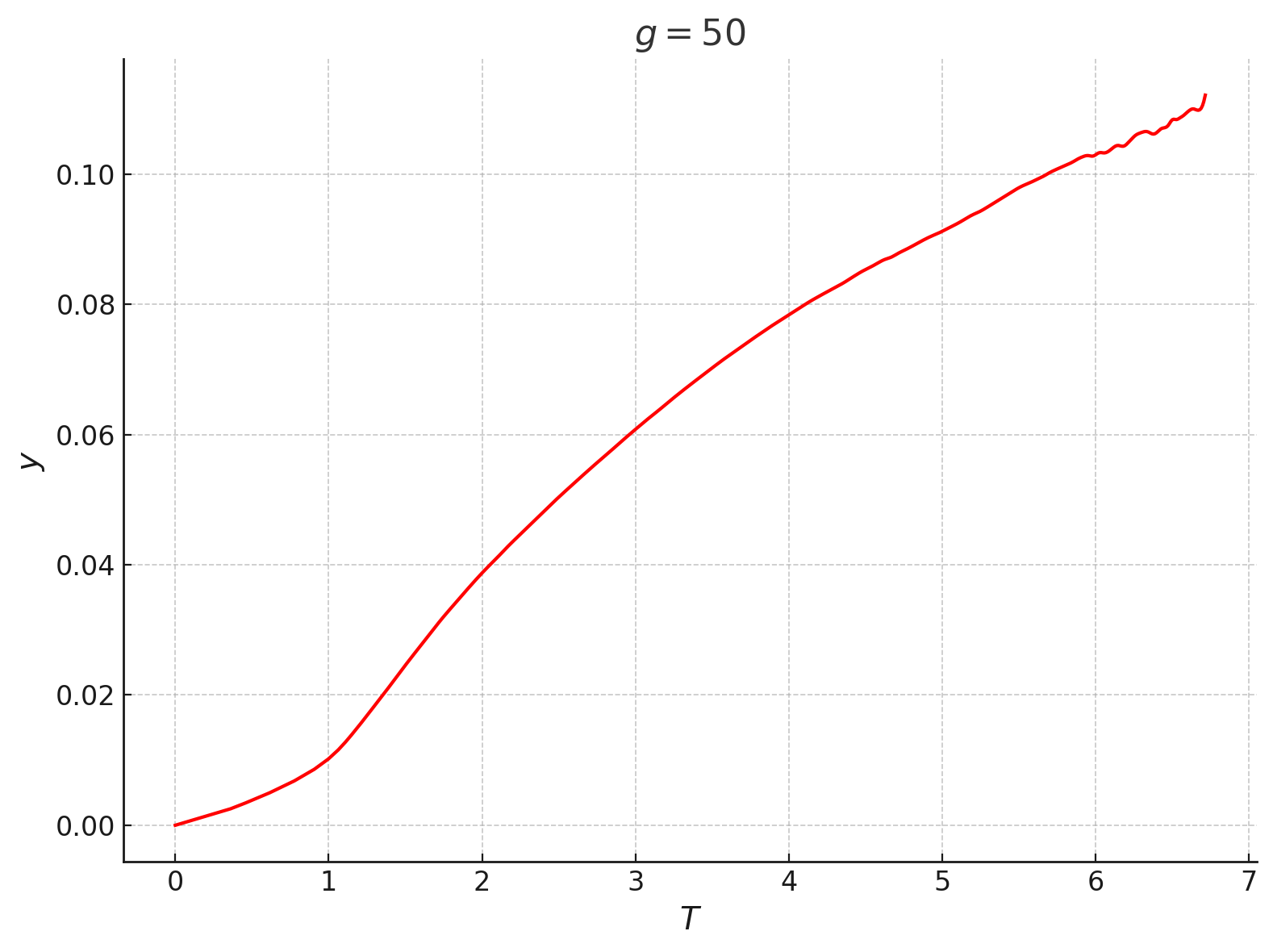}
        \caption{$y$ vs $T$, $g=50$.}
        \label{fig:phi4_T_g50}
    \end{subfigure}
    \caption{Comparison of $E$ vs $y \equiv \phi(M_1 M_2)$, $E$ vs $T$, and $y$ vs $T$ for $g=50$.}
    \label{fig:g50}
\end{figure}

\paragraph{}
We further apply the master field optimization framework to interacting theories, with $g=2, 10, 50$, see \Cref{fig:E_phi4_g2,fig:E_phi4_g10,fig:E_phi4_g50}. This gives the finite temperature expectation value of the loops as in parametric form. The thermal energy is then evaluated as given by the collective potential associated with $H_{+}/2$. To determine the relationship between temperature and $y$ we have several schemes. As the most direct one can use the thermal energy to calibrate the data. The thermal energy $E$ at the corresponding coupling constant $g$ can be evaluated asymptotically. This was given in \cite{Cho:2024kxn} based on  perturbative expansion of the functional integral in the high temperature limit, and based on the long string effective theory in the low temperature limit. Then we create a regression model of thermal energy on temperature, bridging the low-temperature and high-temperature regime, see \Cref{fig:E_T_g2,fig:E_T_g10,fig:E_T_g50}. By comparing $\phi_4$ and temperature through thermal energy, we can establish the relationship between temperature and $\phi_4$, see \Cref{fig:phi4_T_g2,fig:phi4_T_g10,fig:phi4_T_g50}.

\begin{figure}[htb!]
    \centering
    \begin{subfigure}{0.3\textwidth}
        \centering
        \includegraphics[width=\linewidth]{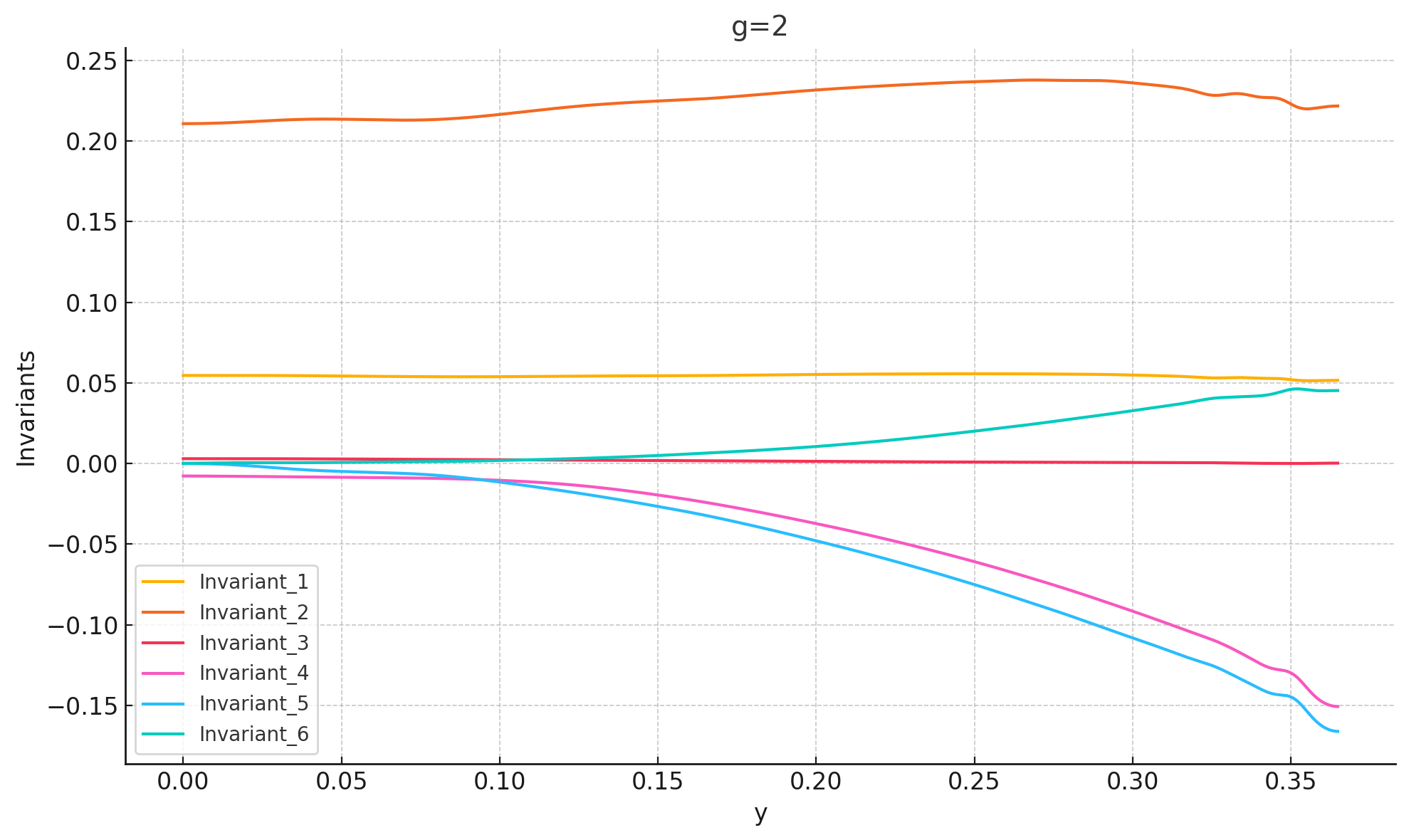}
        \caption{Invariants vs $y$, $g=2$.}
        \label{fig:invts_2}
    \end{subfigure}%
    \hfill
    \begin{subfigure}{0.3\textwidth}
        \centering
        \includegraphics[width=\linewidth]{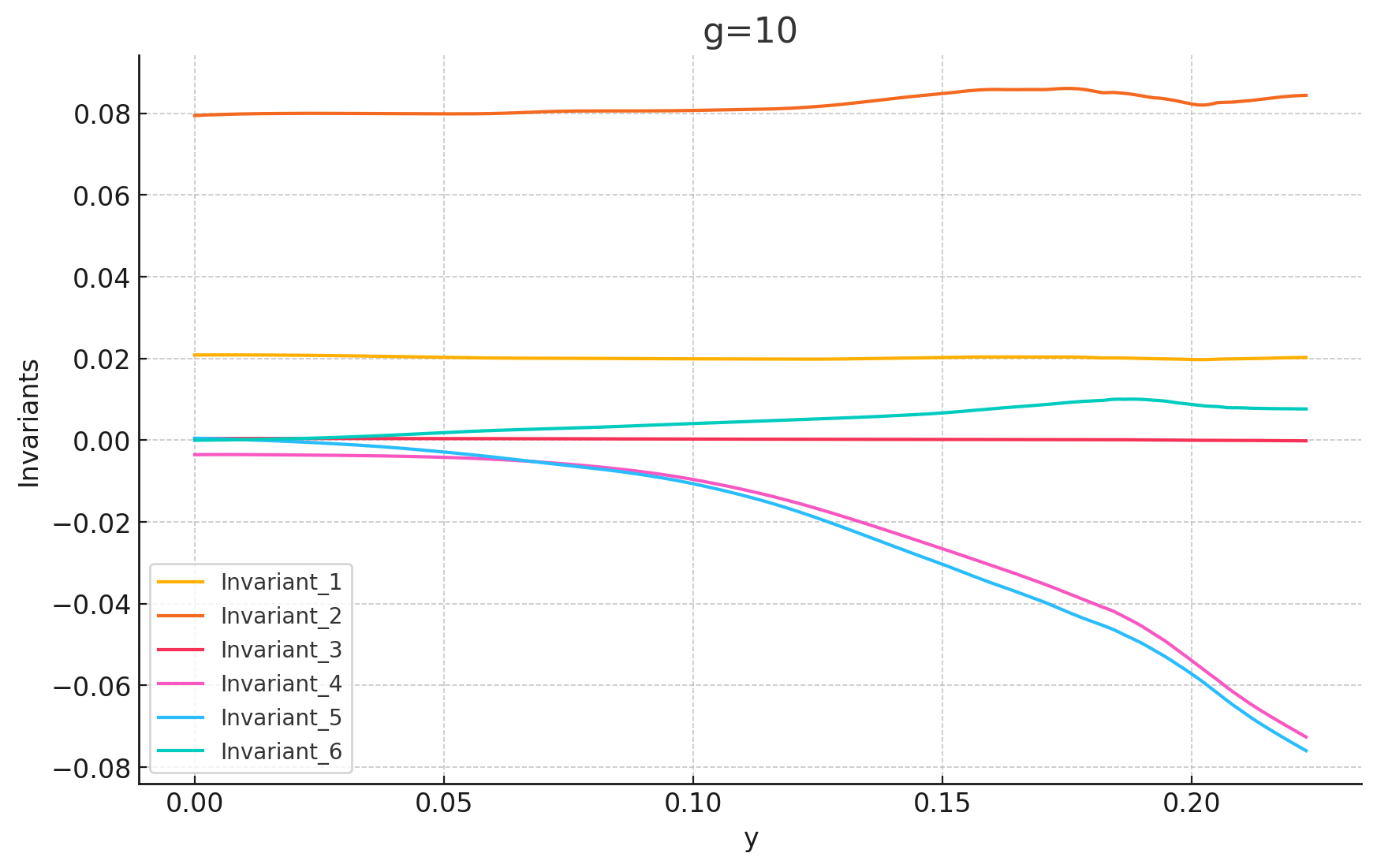}
        \caption{Invariants vs $y$, $g=10$.}
        \label{fig:invts_10}
    \end{subfigure}%
    \hfill
    \begin{subfigure}{0.3\textwidth}
        \centering
        \includegraphics[width=\linewidth]{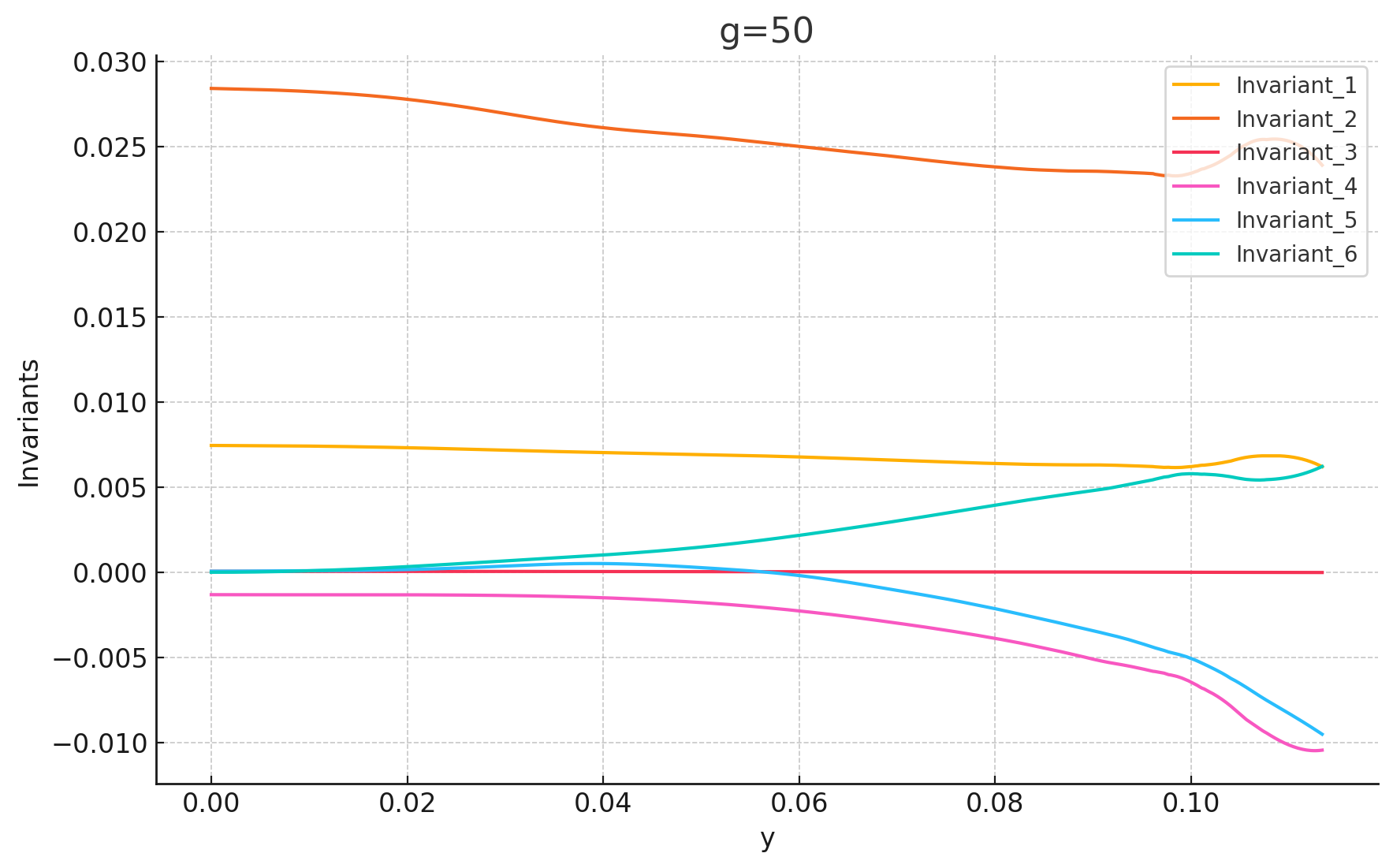}
        \caption{Invariants vs $y$, $g=50$.}
        \label{fig:invts_50}
    \end{subfigure}
    \caption{Comparison of invariants vs $y \equiv \phi(M_1 M_2)$ for $g=2$, $g=10$ and $g=50$.}
    \label{fig:invts}
\end{figure}

\paragraph{}
Once the thermal energy is obtained, thermal loops can also be 
given in terms of the temperature. For example, at $T = 4.772$ and $g = 50$, we tabulate the low lying thermal loop values in \Cref{tab:thermal_loops}. In \Cref{tab:spectrum} we present the low lying spectrum in the decreasing order, where only the first 15 numbers are physical frequencies due to the truncation size $\ell_{\max} = 4$ used in the numerical optimization \cite{Koch:2021yeb}. We also monitor some combinations of thermal loops at finite couplings, which are either invariant in free theory at finite temperature:
\begin{align}
    \text{invariant}_1 &= \phi_3\phi_5 - \phi_4^2 \, , \\
    \text{invariant}_2 &= \phi_{10} + \phi_{15} - 2\phi_{13} \, , \\
    \text{invariant}_3 &= \phi_{12}^2 - \phi_{11}\phi_{14} \, , \\
    \text{invariant}_4 &= \phi_{10} + \phi_{15} - \left(\phi_3 + \phi_5\right)^2 \, , \\
    \text{invariant}_5 &= 2\phi_{13} - \left(2\phi_4\right)^2 \, ,
\end{align}
or invariant in interacting theory at zero temperature:
\begin{equation}
    \text{invariant}_6 = \phi_{12} - \phi_3\phi_5 \, .
\end{equation}
The results are presented in \Cref{fig:invts}. Although we do not have supporting evidence, from pure observation $\text{invariant}_1$ and $\text{invariant}_3$ seem to be preserved against temperature even for the interacting cases.


\begin{table}[htb!]
    \centering
    \begin{small}
    \begin{tabular}{|c|c|c|}
        \hhline{|=|=|=|}
        loop index & loop word & thermal value \\ \hline
        $\phi_0$ & [] & $1.0000 \times 10^{0}$ \\ \hline
        $\phi_1$ & [1] & $1.8025 \times 10^{-2}$ \\ \hline
        $\phi_2$ & [2] & $1.9265 \times 10^{-2}$ \\ \hline
        $\phi_3$ & [11] & $1.1945 \times 10^{-1}$ \\ \hline
        $\phi_4$ & [12] & $8.9698 \times 10^{-2}$ \\ \hline
        $\phi_5$ & [22] & $1.1952 \times 10^{-1}$ \\ \hline
        $\phi_6$ & [111] & $-8.9861 \times 10^{-5}$ \\ \hline
        $\phi_7$ & [112] & $-1.5823 \times 10^{-3}$ \\ \hline
        $\phi_8$ & [122] & $-1.8023 \times 10^{-3}$ \\ \hline
        $\phi_9$ & [222] & $-9.6334 \times 10^{-5}$ \\ \hline
        $\phi_{10}$ & [1111] & $2.5822 \times 10^{-2}$ \\ \hline
        $\phi_{11}$ & [1112] & $1.8397 \times 10^{-2}$ \\ \hline
        $\phi_{12}$ & [1122] & $1.8728 \times 10^{-2}$ \\ \hline
        $\phi_{13}$ & [1212] & $1.3714 \times 10^{-2}$ \\ \hline
        $\phi_{14}$ & [1222] & $1.8400 \times 10^{-2}$ \\ \hline
        $\phi_{15}$ & [2222] & $2.5851 \times 10^{-2}$ \\ \hline
        $\phi_{16}$ & [11111] & $3.6959 \times 10^{-4}$ \\ \hline
        $\phi_{17}$ & [11112] & $8.4678 \times 10^{-5}$ \\ \hline
        $\phi_{18}$ & [11122] & $-1.5357 \times 10^{-4}$ \\ \hline
        $\phi_{19}$ & [11212] & $2.3674 \times 10^{-4}$ \\ \hline
        $\phi_{20}$ & [11222] & $-1.0893 \times 10^{-4}$ \\ \hline
        $\phi_{21}$ & [12122] & $1.9144 \times 10^{-4}$ \\ \hline
        $\phi_{22}$ & [12222] & $3.8191 \times 10^{-5}$ \\ \hline
        $\phi_{23}$ & [22222] & $2.7616 \times 10^{-4}$ \\ \hline
        $\phi_{24}$ & [111111] & $6.3226 \times 10^{-3}$ \\ \hline
        $\phi_{25}$ & [111112] & $4.4043 \times 10^{-3}$ \\ \hline
        $\phi_{26}$ & [111122] & $4.2637 \times 10^{-3}$ \\ \hline
        $\phi_{27}$ & [111212] & $3.1857 \times 10^{-3}$ \\ \hline
        $\phi_{28}$ & [111222] & $3.9538 \times 10^{-3}$ \\ \hline
        $\phi_{29}$ & [112112] & $3.3991 \times 10^{-3}$ \\ \hline
        $\phi_{30}$ & [112122] & $3.3638 \times 10^{-3}$ \\ \hline
        $\phi_{31}$ & [112212] & $3.3638 \times 10^{-3}$ \\ \hline
        $\phi_{32}$ & [112222] & $4.2958 \times 10^{-3}$ \\ \hline
        $\phi_{33}$ & [121212] & $2.4134 \times 10^{-3}$ \\ \hline
        $\phi_{34}$ & [121222] & $3.2131 \times 10^{-3}$ \\ \hline
        $\phi_{35}$ & [122122] & $3.4309 \times 10^{-3}$ \\ \hline
        $\phi_{36}$ & [122222] & $4.4460 \times 10^{-3}$ \\ \hline
        $\phi_{37}$ & [222222] & $6.3568 \times 10^{-3}$ \\ \hline
    \end{tabular}
    \end{small}
    \caption{Thermal loops at $T = 4.772$ and $g = 50$}
    \label{tab:thermal_loops}
\end{table}

\begin{table}[h!]
\centering
\begin{small}
\begin{tabular}{|c|c|}
\hhline{|=|=|}
$i$ & $\varepsilon_i^2$ \\ 
\hline
1 & $7.9048 \times 10^{2}$ \\ \hline
2 & $7.4777 \times 10^{2}$ \\ \hline
3 & $4.6775 \times 10^{2}$ \\ \hline
4 & $4.4722 \times 10^{2}$ \\ \hline
5 & $3.1744 \times 10^{2}$ \\ \hline
6 & $2.4456 \times 10^{2}$ \\ \hline
7 & $2.2308 \times 10^{2}$ \\ \hline
8 & $1.1730 \times 10^{2}$ \\ \hline
9 & $1.3100 \times 10^{2}$ \\ \hline
10 & $5.8919 \times 10^{1}$ \\ \hline
11 & $5.6492 \times 10^{1}$ \\ \hline
12 & $1.5542 \times 10^{1}$ \\ \hline
13 & $9.2255 \times 10^{0}$ \\ \hline
14 & $6.1154 \times 10^{-1}$ \\ \hline
15 & $-5.5608 \times 10^{-1}$ \\ \hline
16 & $1.3167 \times 10^{-1}$ \\ \hline
17 & $-5.4453 \times 10^{-2}$ \\ \hline
18 & $-1.6194 \times 10^{-2}$ \\ \hline
19 & $-1.6194 \times 10^{-2}$ \\ \hline
20 & $1.6081 \times 10^{-2}$ \\ \hline
21 & $8.3959 \times 10^{-3}$ \\ \hline
22 & $8.3959 \times 10^{-3}$ \\ \hline
23 & $-6.5517 \times 10^{-3}$ \\ \hline
24 & $-6.5517 \times 10^{-3}$ \\ \hline
25 & $1.3367 \times 10^{-3}$ \\ \hline
26 & $1.3367 \times 10^{-3}$ \\ \hline
27 & $-4.8562 \times 10^{-3}$ \\ \hline
28 & $1.6919 \times 10^{-3}$ \\ \hline
29 & $1.6919 \times 10^{-3}$ \\ \hline
30 & $-2.5237 \times 10^{-3}$ \\ \hline
31 & $-2.5237 \times 10^{-3}$ \\ \hline
32 & $-2.1308 \times 10^{-3}$ \\ \hline
33 & $-8.6696 \times 10^{-4}$ \\ \hline
34 & $-8.6696 \times 10^{-4}$ \\ \hline
35 & $8.3382 \times 10^{-4}$ \\ \hline
36 & $4.7101 \times 10^{-4}$ \\ \hline
37 & $-1.9595 \times 10^{-4}$ \\ \hline
\end{tabular}
\end{small}
\caption{Spectrum at $T = 4.772$ and $g = 50$}
\label{tab:spectrum}
\end{table}


\clearpage
\newpage

\subsection{KMS conditions}

To summarize, we have seen that in the thermofield scheme the temperature itself appears as a symmetry parameter. The minimization for the thermal energy, the thermal loops, and all correlators is expressed as functions of the parameter, chosen to be $y \equiv \operatorname{tr}(M \tilde{M}) / N^{2}$. The most direct translation to the temperature dependence (instead of $y$) is to independently evaluate $y(T)$ directly, and in the above we have used low and high temperature asymptotics for that.

\paragraph{}
However, as we will now elaborate, there is a solely independent scheme for identification of the temperature, which follows from the KMS relations. Generally, consider the KMS condition on the matrix $M$
\begin{equation}
    \tilde{M} = M\left(\frac{\beta}{2}\right) = \ee^{-\beta H_{+}/2}M \ee^{\beta H_{+}/2} \, .
\end{equation}
At high temperature, we can expand the right-hand-side in terms of $\beta$ as
\begin{equation}
    \tilde{M} = M + \frac{\beta}{2}\Pi - \frac{\beta^2}{8}\left(M + \frac{4g}{N} M^3\right) + \cdots \, .
\end{equation}
The KMS condition can be transferred to loops by acting $\tilde{M}$ on any operator, e.g.,
\begin{equation}
    \tr(M\tilde{M}) = \tr(M^2) + \frac{\beta}{2}\tr(M\Pi) - \frac{\beta^2}{8}\left(\tr(M^2) + \frac{4g}{N}\tr(M^4)\right) + \cdots \, ,
\end{equation}
or equivalently,
\begin{equation}
    \phi_4 = \left(1 - \frac{\beta^2}{8}\right) \phi_3 - \frac{\beta^2g}{2}\phi_{10} - \frac{\beta}{4}\sum_{c, c'}d(c)\phi(c)\Omega_+^{-1}(c,c')\omega_+(c') \, ,
\end{equation}
where $d(c)$ counts the number of $M$'s in loop $c$. Solving this equation for $\beta$, we can determine the temperature from the data explicitly in the extreme limit. We note that the corrections feature the combination $g/T$.

\paragraph{}
Proceeding to two-point correlators we next show that an exact set of KMS relations will be  operational at $O(1)$ in $1/N$-expansion, which can then be used for measuring the temperature. The thermofield potential can also be written as
\begin{equation}
    \hat{V}_-[\phi] = \frac{1}{8}\omega_-^{\rm{T}}\Omega_+^{-1}\omega_+ + \frac{1}{2}(\phi_3-\phi_5) + g(\phi_{10} - \phi_{15}) \, .
\end{equation}
Consider the small kernel fluctuation around the thermal solution with
\begin{equation}
    \hat{L}^{(2)} = \frac{1}{2}\dot{\eta}^{\rm{T}}\Omega^{-1}_{-, f}\dot{\eta} - \frac{1}{2}\eta^{\rm{T}} V''_{-, f} \eta = -\frac{1}{2} \eta^{\rm{T}}\left(\Omega^{-1}_{-, f}\partial_t^2 + V''_{-, f}\right)\eta\, ,
\end{equation}
where $\phi=\phi_f + \eta / N$ and $f$ is a free parameter due to the existence of the dynamical symmetry. The mass matrix takes the form
\begin{equation}
    V''_- = \frac{1}{8}\left(\omega''_- \Omega_+^{-1}\omega_+ + \omega_-\Omega_+^{-1}\omega''_+ -2 \omega'_-\Lambda_+\omega_+ -2 \omega_-\Lambda_+\omega'_+ +2 \omega'_-\Omega_+^{-1}\omega'_+ + 2 \omega_- \Lambda_+\Omega_+\Lambda_+\omega_+\right) \, ,
\end{equation}
where we have introduced
\begin{equation}
    \Lambda_+ = \Omega_+^{-1}\Omega'_+\Omega_+^{-1} \, .
\end{equation}
We need to solve the eigenvalue equation
\begin{equation}
    \left(\omega_\alpha^2 - \Omega_{-, f}V''_{-, f}\right)W_\alpha = 0 \, ,
\end{equation}
where $\alpha = \{i, \sigma\}$, $i = 0, 1, 2, \cdots$, $\sigma = \pm$. It is obvious that each $\omega_\alpha$ has a two-fold degeneracy with a pair of eigenvectors $W_{\alpha}$, except for the zero mode $\omega_0$ with zero eigenvector $W_0$. We can find $W$ such that
\begin{equation}
    W^{{T}}_{\alpha}\Omega_{-, f}^{-1} W_{\alpha'} = m_{\alpha}\delta_{\alpha,\alpha'} \, .
\end{equation}
In the normal mode basis
\begin{equation}
    \eta = W q \, ,
\end{equation}
the Lagrangian can be diagonalized as
\begin{equation}
    \hat{L}^{(2)} = \frac{1}{2}\sum_{\alpha} m_{\alpha} \dot{q}_{\alpha}^2 - \frac{1}{2}\sum_{\alpha} m_{\alpha}\omega_\alpha^2 q_{\alpha}^2 = -\frac{1}{2}\sum_{\alpha} m_{\alpha}q_{\alpha}^{\rm{T}}\left(\partial_t^2 + \omega_\alpha^2\right)q_{\alpha} \, .
\end{equation}
The normal mode propagators associated with eigenvalue $\omega_\alpha$ in the Fourier space are given by
\begin{equation}
    d_\alpha(E) = \frac{1}{m_{\alpha}} \frac{1}{E^2 - \omega_\alpha^2 + \ii \sigma \epsilon} \, ,
\end{equation}
or in coordinate space
\begin{equation}
    d_\alpha(t) = \frac{\ee^{-\ii\omega_\alpha |t|}}{2m_{\alpha}\omega_\alpha} \, .
\end{equation}
The zero mode propagator is simply a non-zero finite constant
\begin{equation}
    d_0 = \frac{1}{2m_0\omega_0} \, .
\end{equation}
\begin{figure}[tbh!]
    \begin{center}
        \includegraphics[width=0.5\textwidth]{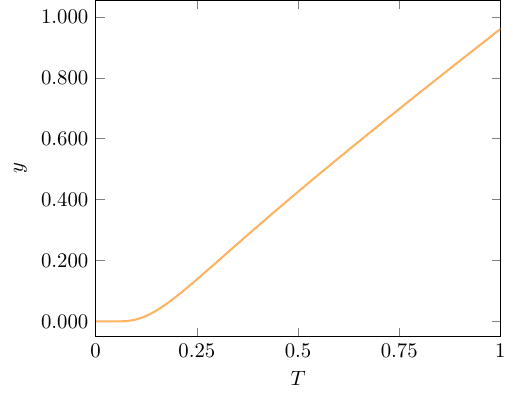}
        \caption{$y \equiv \phi(M \tilde{M})$ versus $T$ for free theory case.}
        \label{fig:y_vs_T_free_theory}
    \end{center}
\end{figure}
We can assemble the small fluctuation propagators in the normal mode basis
\begin{equation}
    \ii D_0(t) = \braket{q(t)q^{\rm{T}}(0)} = \ii \operatorname{diag}\left(d_{n}, \cdots, d_{1}, d_0, d_{-1}, \cdots, d_{-n}\right) \, ,
\end{equation}
and $W$ should be arranged accordingly. Performing $W$ transformation, we obtain the thermal propagator
\begin{equation}
    \ii D_f(t) = \braket{\eta(t)\eta^{\rm{T}}(0)} = \ii W D_0(t) W^{\rm{T}} \, .
\end{equation}
Recall that KMS conditions require
\begin{equation}
    M(t - \ii \beta/2) = \Tilde{M}(t) \, , \quad \Tilde{M}(t - \ii \beta/2) = M(t) \, ,
\end{equation}
which lead to a series of conditions on the thermal propagators, e.g., 
\begin{equation}
\begin{split}
    &(D_f)_{1,1}(t) = (D_f)_{2,1}(t - \ii\beta/2)\, , \quad (D_f)_{2,2}(t) = (D_f)_{1,2}(t - \ii\beta/2) \, , \\
    &(D_f)_{3,3}(t) = (D_f)_{5,3}(t - \ii\beta/2)\, , \quad (D_f)_{5,5}(t) = (D_f)_{3,5}(t - \ii\beta/2) \, .
\end{split}
\end{equation}
To be more specific, let us take $t \rightarrow 0^+$ without loss of generality. The equations above read
\begin{equation}
\begin{split}
    &\sum_{\alpha} \frac{1}{2 m_\alpha \omega_\alpha} W^2_{1\alpha} = \sum_{\alpha} \frac{\ee^{-\omega_\alpha \beta/2}}{2 m_\alpha \omega_\alpha} W_{1\alpha}W_{2\alpha} =
    \sum_{\alpha} \frac{1}{2 m_\alpha \omega_\alpha} W^2_{2\alpha} \, , \\
    &\sum_{\alpha} \frac{1}{2 m_\alpha \omega_\alpha} W^2_{3\alpha} = \sum_{\alpha} \frac{\ee^{-\omega_\alpha \beta/2}}{2 m_\alpha \omega_\alpha} W_{3\alpha}W_{5\alpha} =
    \sum_{\alpha} \frac{1}{2 m_\alpha \omega_\alpha} W^2_{5\alpha} \, .
\end{split}
\end{equation}
The above KMS relations featured for $O(1)$ two-point correlators of arbitrary loops determine the inverse temperature $\beta$ from the $O(1)$ data. For the free theory, where such data was found in \Cref{sec:Free_theory} explicitly, the above equations can be verified to hold, and give
\begin{equation}
    \operatorname{sech} \frac{\beta}{2} =\tanh 2f \, .
\end{equation}
We then have the plot of the $y-T$ relation in \Cref{fig:y_vs_T_free_theory}. A more general evaluation of $O(1)$ correlation functions and a study of 
the corresponding KMS relations in loop space will be given in 
future work.

\section{Conclusion}
\label{sec:Conclusion}

There are a number of closely related topics that we did not have space to consider. At the free level the transformation properties of thermal loops also allow evaluation of multi-point correlation functions. Likewise is an extension of the Schur basis \cite{deMelloKoch:2007nbd,deMelloKoch:2007rqf,Bhattacharyya:2008rb} to the thermal case. The use of $H_+$ and its thermal deformation along the lines discussed in \Cref{sec:Free_theory} can also be pursued and studied at the interacting level. Of some interest would be a comparison with the well-known deformation of \cite{Maldacena:2018lmt}. The most interesting and most relevant would be a further detailed study of the KMS equations, presented in the context of two-point correlators in \Cref{sec:Interacting_theory}. These were shown to be given in terms of the data, found in the numerical solution. The relations therefore provide an extensive set of consistency relations, in particular on the symmetry properties of the thermal solution. Related to the correlators would be a concrete construction of the thermofield state functional itself. At Large $N$ it takes a Gaussian form and can again be given in terms of the numerical data. This and other interesting studies will be addressed in the future.

\section*{Acknowledgements}
\label{sec:Acknowledgements}
We are thankful for many discussions with colleagues on the topics presented. We acknowledge collaboration and detailed discussions with Robert de Mello Koch, Jo\~{a}o Rodrigues, Junggi Yoon, Sumit Das and Minjae Cho. AJ would like to acknowledge the hospitality of PCTP in Princeton and the Simmons Collaboration on QCD and Quark Confinement. Discussions with David Gross, Igor Klebanov and Larry Yaffe are also acknowledged. The work of A.J. and J.Z. is supported by the U.S. Department of Energy under contract DE-SC0010010. XL is supported by the National Natural Science Foundation of China (NSFC): Individual Grant No. 12374138, Research Fund for International Senior Scientists No. 12350710180, and National Key R\&D Program of China (Project ID: 2019YFA0308603).

\appendix

\section{Loop space and loop functions}
\label{appendix:loop_space_and_loop_functions}

A systematic study of large $N$ limit generally requires a change of representation from the original matrix valued variables to invariant variables, which we refer to as ``loops''. In this section, we present a brief review of the loop space representation and introduce some basic terminologies. For more technical discussions, one may refer to \cite{Koch:2021yeb,Mathaba:2023non}. Simply speaking loop variables are the traces of various matrix products. As an example, for the case of two hermitian matrices we have
\begin{equation} 
\phi (C) = \frac{1}{N^{\frac{\len(C)}{2}+1}} \operatorname{tr}(M_1^{n_1}M_2^{n_2}M_1^{n'_1}M_2^{n_2'}\cdots) \, .
\end{equation}
Here $C$ represents the \emph{word} constructed from the alphabet of the matrices $\{M_1, M_2\}$. $\len(C)$ denotes the length of the word $C$. Loops are invariant under the global U($N$) transformation $M_i \rightarrow U^\dagger M_i U$, $i=1,2$ with $U^\dagger U = 1$. The word specifies the order in which matrices are multiplied before taking the trace. The invariant loop variables are then described by all of the words with cycling identification. For example, $\tr(M_1 M_1 M_2)$ is equal to $\tr(M_1 M_2 M_1)$ and also $\tr(M_2 M_1 M_1)$ due to the cyclicity of trace, hence they all refer to the same invariant loop variable. Removing this redundancy due to the cyclicity, we still have an infinite number of loops, which span an infinite dimensional linear space. We call this space the \emph{loop space}, denoted $V$. Let $V_{\ell}$ denote the subspace which consists of all (inequivalent) loops of length $\ell$, the loop space can be written as a direct sum of these subspaces
\begin{equation}
  V = \bigoplus_{\ell=0}^{\infty} V_{\ell} \, .
\end{equation}

The first question to ask is: for a given $\ell$, what is the dimension of $V_{\ell}$? This counting problem can be solved by applying Polya theory. Denote $\varphi(d)$ the Euler totient function, we have the single trace partition function 
\begin{align}
F(x,y) = & \sum_{n}\sum_{n|d}\frac{\varphi (d)}{n}(x^d + y^d)^{\frac{n}{d}} \nonumber \\
= & \sum_{n,m=1}^\infty {\cal N}_{n,m}x^n y^m \, . \label{eq:dim_Vl_generating_function}
\end{align}
The degree $\ell$ contribution to $F(x,y)$ then counts independent loops constructed from $\ell$ matrices. Then the dimension of $V_{\ell}$ is given by
\begin{equation}
    \dim(V_{\ell}) = \sum_{n + m = \ell} \mathcal{N}_{n,m} 
    = \sum_{d \mid \ell} \frac{1}{\ell} \, 2^{\frac{\ell}{d}} \, \varphi(d) \, .
\end{equation}

The dimensions of $V_{\ell}$ for $\ell \leq 30$ are explicitly presented in \Cref{tab:dim_V_l}. As shown in the table, the dimensions increases rapidly when $\ell$ increases. For notational convenience, we may also assign an integer index to each loop. Our conventions for the first 16 loops of the two-matrix systems are exhibited in \Cref{tab:loopinfo2}.

\paragraph{}
For systems involving $s$ matrices the generating function \eqref{eq:dim_Vl_generating_function} is modified to be
\begin{equation}
  F(x_1, x_2, \dots, x_s) = \sum_{n} \sum_{n|d} \frac{\varphi(d)}{n} Z_1(x_1^d, x_2^d, \dots, x_s^d)^{\frac{n}{d}} \, , 
\end{equation}
where $Z_1(x_1, x_2, \dots, x_s) = x_1 + x_2 + \dots x_s$. Then $\dim(V_l)$ for $s$-matrix systems can be read off similarly from the corresponding blind function $F(\alpha, \alpha, \dots, \alpha)$.

\begin{table}[htb!]
\begin{center}
    \begin{tabular}{||c|c|c|c|c|c|c|c|c|c|c|c|c|c|c|c||}
    \hline
    $\ell$ & 0 & 1 & 2 & 3 & 4 & 5 & 6 & 7 & 8 & 9 & 10 & 11 & 12 & 13 & 14\\
    \hline
    $\operatorname{dim}(V_{\ell})$ & 1 & 2 & 3 & 4 & 6 & 8 & 14 & 20 & 36 & 60 & 108 & 188 & 352 & 632 & 1182 \\
    \hline
    \end{tabular}
    \begin{tabular}{||c|c|c|c|c|c|c|c|c|c||}
    \multicolumn{4}{c}{\vspace{0.2cm}}\\
    \hline
    $\ell$ & 15 & 16 & 17 & 18 & 19 & 20 & 21 & 22 & 23 \\
    \hline
    $\operatorname{dim}(V_{\ell})$ & 2192 & 4116 & 7712 & 14602 & 27596 & 52488 & 99880 & 190746 & 364724 \\
    \hline
    \end{tabular}
    \begin{tabular}{||c|c|c|c|c|c|c|c||}
        \multicolumn{4}{c}{\vspace{0.2cm}}\\
        \hline
        $\ell$ & 24 & 25 & 26 & 27 & 28 & 29 & 30 \\
        \hline
        $\operatorname{dim}(V_{\ell})$ & 699252 & 1342184 & 2581428 & 4971068 & 9587580 & 18512792 & 35792568 \\
        \hline
        \end{tabular}
  \caption{Dimensions of loop subspace $V_{\ell}$ for two-matrix systems}
  \label{tab:dim_V_l}
\end{center}
\end{table}

\begin{figure}[t!]
\begin{center}
\includegraphics{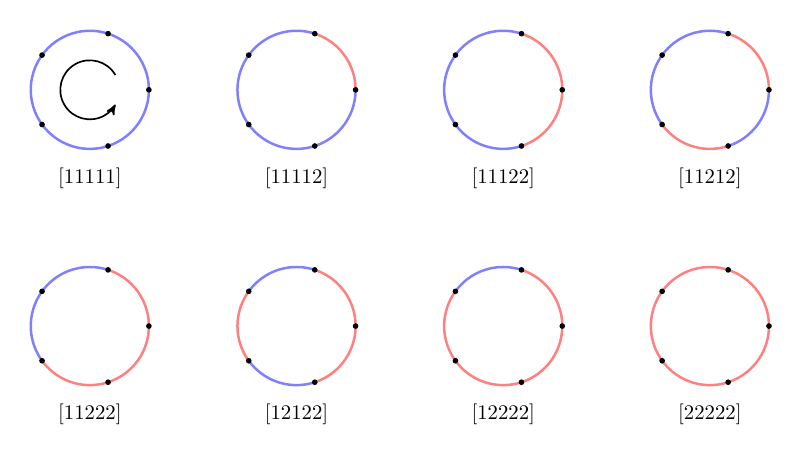}
\caption{Inequivalent loops of lengths 5.}
\label{fig:loops_l5}
\end{center}
\end{figure}

\paragraph{}
Let us then give the geometrical representations of the loops. Since we are only interested in the singlets, it is enough to adopt the one-line notion in stead of the double-line representation in the context of Feynman diagrams. We can use different colored arcs for different matrices. The loop of length $\ell$ then is represented as a oriented circle with $\ell$ colored arcs. The cyclicity of the matrix trace is reflected in the fact that circles can be rotated along their center axis with angles $2 \pi n / \ell$ ($0 \leq n \leq \ell-1$), exhibiting the $\mathbb{Z}_\ell$ symmetry. In \Cref{fig:loops_l5} we present an example for all loops in $V_{5}$, where for $M_1$ we assign blue arcs and for $M_2$ we assign red arcs. Also note that here we use the convention that the loops are read off in the counter-clockwise direction of the corresponding circles. Reading the arcs in the opposite directions is equivalent to reverse the words $C$ and hence produces the hermitian conjugate of the loops for hermitian matrix models.

\begin{figure}[htb!]
    \begin{center}
      
    \includegraphics[width=0.9999\textwidth]{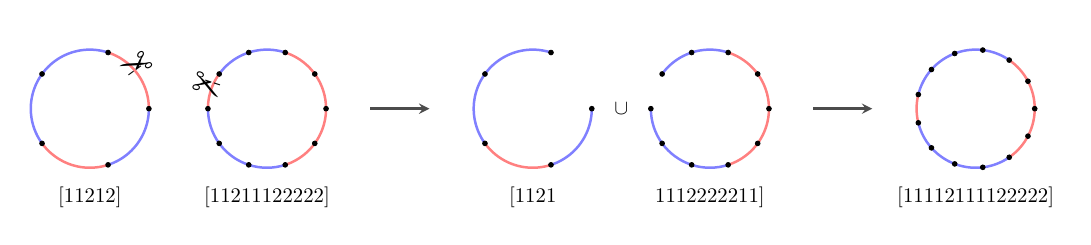}
    
    \caption{Illustration of a contribution for the loop joining process in $\Omega_{22}$.}
    \label{fig:loop_joining_illustration}
    \end{center}
    \end{figure}

\paragraph{}
With the conception of loops we can then consider two classes of important functions of loops that are frequently encountered in the loop space representations. We again focus on the two-matrix case. The first class is \emph{loop joining}. Let us consider the function
\begin{equation} \label{eq:Omega_ij}
    \Omega_{ij}(C_1, C_2) = N^2 \tr(\pdv{\phi(C_1)}{M_i} \pdv{\phi(C_2)}{M_j}) \, , \quad i, j = 1, 2 \, .
\end{equation}
A geometric illustration for one contribution of $\Omega_{22}(M_1^2 M_2 M_1 M_2, \, M_1^2 M_2 M_1^3 M_2^5)$ is presented in \Cref{fig:loop_joining_illustration}. The derivatives of $\phi(C)$ with respect to $M_i$ are represented by cutting the arc corresponding to $M_i$ in $\phi(C)$, resulting in two open arcs. The trace operation is represented by gluing the two open arcs such that a closed loop is formed. The loop joining is then a summation of all such operations. As a result, $\Omega_{ij}$ takes two loops and produces a linear function of loops:
\begin{equation}
    \Omega_{ij}(C_1, C_2) = \sum_{C} \, \mathsf{j}_{i j}(C_1, C_2; C) \, \phi(C) \, ,
\end{equation}
where $\len(C) = \len(C_1) + \len(C_2) - 2$ and $\mathsf{j}_{ij}(C_1, C_2; C)$ some integers.

\begin{figure}[htb!]
    \begin{center}
      
    \includegraphics[width=0.9999\textwidth]{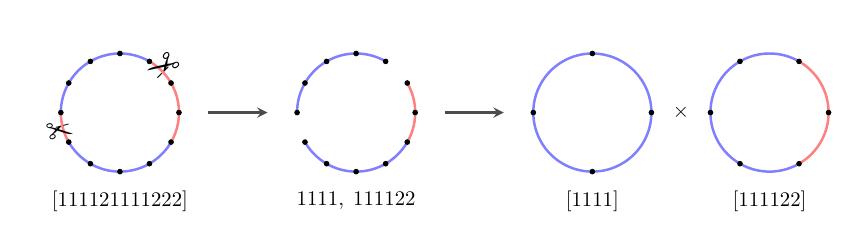}
    
    \caption{Illustration of a contribution for the loop splitting process in $\omega_{22}$.}
    \label{fig:loop_splitting_illustration}
    \end{center}
    \end{figure}

\paragraph{}
The other class is \emph{loop splitting}, representing the opposite operation of the loop joining. Let us consider the function
\begin{equation} \label{eq:omega_ij}
    \omega_{ij}(C) = \tr(\pdv[2]{\phi(C)}{M_i}{M_j}) \, , \quad i, j = 1, 2 \, .
\end{equation}
A geometric illustration for one contribution of $\omega_{22}(M_1^4 M_2 M_1^4 M_2^3)$ is presented in \Cref{fig:loop_splitting_illustration}. The two derivatives of $\phi(C)$ with respect to $M_i$ and $M_j$ are represented by cutting the corresponding arcs in $\phi(C)$, resulting in two disconnected open arcs. The trace operation then is represented by separately gluing the endpoints of the two open arcs, and in the end one has a product of two loops. The loop splitting is then a summation of all such operations. Consequently $\omega_{ij}$ takes one loop and produces a quadratic function of loops:
\begin{equation}
    \omega_{ij}(C) = \sum_{(C_1, C_2)} \, \mathsf{p}_{i j}(C; C_1, C_2) \, \phi(C_1) \, \phi(C_2) \, ,
\end{equation}
where $\len(C_1) + \len(C_2) + 2 = \len(C)$ and $\mathsf{p}_{i j}(C; C_1, C_2)$ are some integers.

%

\section{Thermal loop values}
\label{appendix:thermal_loop_values}

\begin{figure}[htb!]
\begin{center}
    \includegraphics[width=0.8\textwidth]{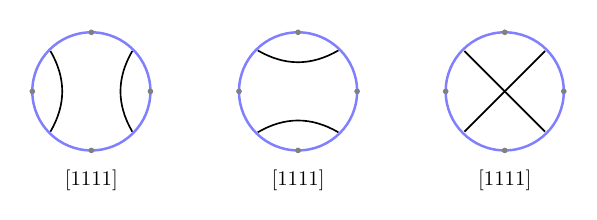}
    \caption{Diagrammatic representation of the Wick contractions for $\langle \tr(M^4) \rangle / N^3$. They are given by chord diagrams. Only the diagrams without intersections (planar diagrams) contribute in the large $N$ limit. The diagrams with intersections, such as the last one (non-planar diagram), are of sub-leading order in $1/N$, and hence vanish in the large $N$ limit.}
    \label{fig:chord_diagrams_tr_M_to_4}
\end{center}
\end{figure}

In this section we present the loop values in free theory, both at zero temperature and finite temperature. Since our analytical strategy presented in \Cref{sec:Free_theory} for computing loops at finite temperature involving the prerequisite knowledge of the loops at zero temperature, we first present a simple method for computing zero temperature loops. We will show that they are given by the number of chord diagrams without intersections, equivalent to the summation of planar diagrams. Let us first consider loops involving only one matrix, namely $\tr(M^n) / N^{n/2 + 1}$. For free theory we simply have 
\begin{equation}
    \frac{1}{N^2} \bra{0} \tr(M^2) \ket{0} = \frac{1}{2} \, .
\end{equation}
The expectation values of higher loops are given by the Wick contractions, and taking the large $N$ limits is equivalent to summing all the Wick contractions without intersections. For example,
\begin{align} \label{eq:tr_M_to_4_Wick_constraction}
    \frac{1}{N^3} \langle \tr(M^4) \rangle & = \frac{1}{N^3} \langle \operatorname{tr}(\wick{\c M \c M} \wick{\c M \c M}) \rangle +  \frac{1}{N^3} \langle \operatorname{tr}(\wick{\c2 M \c1 M \c1 M \c2 M}) \rangle +  \frac{1}{N^3} \langle \operatorname{tr}(\wick{\c1 M\c2 M \c1 M \c2 M}) \rangle \\[2pt]
    & = \frac{1}{4} + \frac{1}{4} + \frac{1}{4 N} \\[2pt]
    & = \frac{1}{2} \, .
\end{align}
We note that the last Wick contraction involves the crossing of the matrices, and hence produces a contribution of sub-leading order in $1/N$. Diagrammatically, we can represent a loop of length $\ell$ as a circle consists of $\ell$ points. Wick contractions then correspond to connecting the points with chords. If there are intersections of the chords, then the corresponding Wick contraction will be zero in the large $N$ limit. Therefore, the loop values at zero temperature are given by the number of chord diagrams without intersections. For example, the chord diagrams corresponding to Wick contractions of \eqref{eq:tr_M_to_4_Wick_constraction} are shown in \Cref{fig:chord_diagrams_tr_M_to_4}. 

\begin{figure}[htb!]
    \begin{center}
        \includegraphics[width=0.8\textwidth]{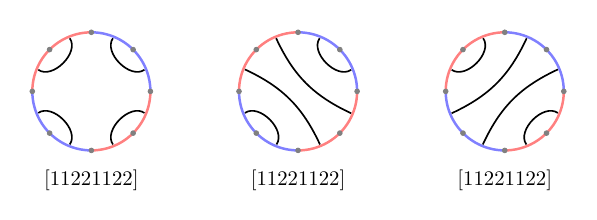}
        \caption{Chord diagrams without intersections corresponding to the loop $\langle \tr(M_1^2 M_2^2 M_1^2 M_2^2) / N^5 \rangle$.}
        \label{fig:chord_diagrams_tr_11221122}
    \end{center}
    \end{figure}

\paragraph{}
This can be straightforwardly generalized to loops involving multiple matrices: different matrices are represented by different colored arcs, and Wick contractions correspond to adding chords of the arcs with same colors. Again, one only sums the non-intersecting chord diagrams. In conclusion, for a given loop $\phi(C)$, we have
\begin{equation}
    \langle \phi(C) \rangle = \frac{1}{2^{\frac{\len(C)}{2}}} \times (\text{\# of chord diagrams connecting same colored arcs without intersections}) \, .
\end{equation}
To give a simple example, consider $\langle \tr(M_1^2 M_2^2 M_1^2 M_2^2) / N^5 \rangle$. We have three chord diagrams without interactions, as illustrated in \Cref{fig:chord_diagrams_tr_11221122}, so the loop value is $3 \times 2^{-4} = 3/16$.

\paragraph{}
The conclusion hints that we can use an iterative algorithm to compute higher loop values at zero temperature, once we know the lower loop values. In addition, the diagrammatic representations illustrate the relation between the loop expectation values and loop splitting. For one-matrix free case we simply have
\begin{equation}
    \braket{\omega_{n}} = n \sum_{k=0}^{n-2} \braket{\phi_k} \braket{\phi_{n-k-2}} = 2 \, n \, \braket{\phi_n} \, .
\end{equation}

\paragraph{}
Loop values at finite temperature in free theory case can be computed via the $G$-transformations discussed in \Cref{sec:Free_theory}. In \Cref{tab:loop_info_V6,tab:loop_info_V8,tab:loop_info_V8_continued} we present the non-zero loop values in $V_6$ and $V_8$ at both zero temperature and finite temperature.

\begin{table}[htb!]
\begin{center}
    \begin{NiceTabular}{|c|c|c|}
    \hhline{|=|=|=|}
    loop word & $\bra{0} \phi^{a} \ket{0}$ & $\bra{0(\beta)} \phi^{a} \ket{0(\beta)}$ \\
    \hline
    \text{[111111]} & $\frac{5}{8}$      & $\frac{5}{8} \cosh^3(2 \theta )$ \\
    \hline
    \text{[111112]} & $0$ & $\frac{5}{8} \sinh (2 \theta ) \cosh^2(2 \theta )$ \\
    \hline
    \text{[111122]} & $\frac{1}{4}$ & $\frac{1}{32} (3 \cosh (2 \theta )+5 \cosh (6 \theta ))$ \\
    \hline
    \text{[111212]} & $0$ & $\frac{5}{16} \sinh (2 \theta ) \sinh (4 \theta )$ \\
    \hline
    \text{[111222]} & $0$ & $\frac{1}{32} (\sinh (2 \theta )+5 \sinh (6 \theta ))$ \\
    \hline
    \text{[112112]} & $\frac{1}{8}$ & $\frac{1}{32} (5 \cosh (6 \theta )-\cosh (2 \theta ))$ \\
    \hline
    \text{[112122]} & $0$ & $\frac{1}{32} (5 \sinh (6 \theta )-3 \sinh (2 \theta ))$ \\
    \hline
    \text{[112212]} & $0$ & $\frac{1}{32} (5 \sinh (6 \theta )-3 \sinh (2 \theta ))$ \\
    \hline
    \text{[112222]} & $\frac{1}{4}$ & $\frac{1}{32} (3 \cosh (2 \theta )+5 \cosh (6 \theta ))$ \\
    \hline
    \text{[121212]} & $0$ & $5 \sinh ^3(\theta ) \cosh ^3(\theta )$ \\
    \hline
    \text{[121222]} & $0$ & $\frac{5}{16} \sinh (2 \theta ) \sinh (4 \theta )$ \\
    \hline
    \text{[122122]} & $\frac{1}{8}$ & $\frac{1}{32} (5 \cosh (6 \theta )-\cosh (2 \theta ))$ \\
    \hline
    \text{[122222]} & $0$ & $\frac{5}{8} \sinh (2 \theta ) \cosh ^2(2 \theta )$ \\
    \hline
    \text{[222222]} & $\frac{5}{8}$ & $\frac{5}{8} \cosh ^3(2 \theta )$ \\
    \hline
    \end{NiceTabular}
    \caption{Expectation values of loops in $V_{6}$ at zero temperature and finite temperature.}
    \label{tab:loop_info_V6}
\end{center}
\end{table}

\begin{table}[htb!]
    \begin{center}
        \begin{NiceTabular}{|c|c|c|}
        \hhline{|=|=|=|}
        loop word & $\bra{0} \phi^{a} \ket{0}$ & $\bra{0(\beta)} \phi^{a} \ket{0(\beta)}$ \\
        \hline
        \text{[11111111]} & $\frac{7}{8}$ & $\frac{7}{8} \cosh ^4(2 \theta )$ \\
        \hline
        \text{[11111112]} & $0$ & $\frac{7}{8} \sinh (2 \theta ) \cosh ^3(2 \theta )$ \\
        \hline
        \text{[11111122]} & $\frac{5}{16}$ & $\frac{1}{16} \cosh ^2(2 \theta ) (7 \cosh (4 \theta )-2)$ \\
        \hline
        \text{[11111212]} & $0$ & $\frac{7}{32} \sinh ^2(4 \theta )$ \\
        \hline
        \text{[11111222]} & $0$ & $\frac{1}{64} (6 \sinh (4 \theta )+7 \sinh (8 \theta ))$ \\
        \hline
        \text{[11112112]} & $\frac{1}{8}$ & $\frac{1}{16} \cosh ^2(2 \theta ) (7 \cosh (4 \theta )-5)$ \\
        \hline
        \text{[11112122]} & $0$ & $\frac{7}{64} \sinh (8 \theta )$ \\
        \hline
        \text{[11112212]} & $0$ & $\frac{7}{64} \sinh (8 \theta )$ \\
        \hline
        \text{[11112222]} & $\frac{1}{4}$ & $\frac{1}{64} (6 \cosh (4 \theta )+7 \cosh (8 \theta )+3)$ \\
        \hline
        \text{[11121112]} & $0$ & $\frac{7}{32} \sinh ^2(4 \theta )$ \\
        \hline
        \text{[11121122]} & $0$ & $\frac{7}{64} \sinh (8 \theta )$ \\
        \hline
        \text{[11121212]} & $0$ & $\frac{7}{8} \sinh ^3(2 \theta ) \cosh (2 \theta )$ \\
        \hline
        \text{[11121222]} & $0$ & $\frac{1}{16} \sinh ^2(2 \theta ) (7 \cosh (4 \theta )+5)$ \\
        \hline
        \text{[11122112]} & $0$ & $\frac{7}{64} \sinh (8 \theta )$ \\
        \hline
        \text{[11122122]} & $\frac{1}{8}$ & $\frac{1}{64} (7 \cosh (8 \theta )+1)$ \\
        \hline
        \text{[11122212]} & $0$ & $\frac{1}{16} \sinh ^2(2 \theta ) (7 \cosh (4 \theta )+5)$ \\
        \hline
        \text{[11122222]} & $0$ & $\frac{1}{64} (6 \sinh (4 \theta )+7 \sinh (8 \theta ))$ \\
        \hline
        \text{[11211212]} & $0$ & $\frac{1}{64} (7 \sinh (8 \theta )-6 \sinh (4 \theta ))$ \\
        \hline
        \text{[11211222]} & $\frac{1}{8}$ & $\frac{1}{64} (7 \cosh (8 \theta )+1)$ \\
        \hline
        \text{[11212122]} & $0$ & $\frac{1}{16} \sinh ^2(2 \theta ) (7 \cosh (4 \theta )+2)$ \\
        \hline
        \text{[11212212]} & $\frac{1}{16}$ & $\frac{1}{64} (-6 \cosh (4 \theta )+7 \cosh (8 \theta )+3)$ \\
        \hline
        \text{[11212222]} & $0$ & $\frac{7}{64} \sinh (8 \theta )$ \\
        \hline
        \text{[11221122]} & $\frac{3}{16}$ & $\frac{1}{64} (7 \cosh (8 \theta )+5)$ \\
        \hline
        \end{NiceTabular}
        \caption{Expectation values of loops in $V_{8}$ at zero temperature and finite temperature.}
        \label{tab:loop_info_V8}
\end{center}
\end{table}

\begin{table}[htb!]
\ContinuedFloat
\begin{center}
    \begin{NiceTabular}{|c|c|c|}
        \hhline{|=|=|=|}
        loop word & $\bra{0} \phi^{a} \ket{0}$ & $\bra{0(\beta)} \phi^{a} \ket{0(\beta)}$ \\
        \hline
        \text{[11221212]} & $0$ & $\frac{1}{16} \sinh ^2(2 \theta ) (7 \cosh (4 \theta )+2)$ \\
        \hline
        \text{[11221222]} & $0$ & $\frac{7}{64} \sinh (8 \theta )$ \\
        \hline
        \text{[11222122]} & $0$ & $\frac{7}{64} \sinh (8 \theta )$ \\
        \hline
        \text{[11222212]} & $0$ & $\frac{7}{64} \sinh (8 \theta )$ \\
        \hline
        \text{[11222222]} & $\frac{5}{16}$ & $\frac{1}{16} \cosh ^2(2 \theta ) (7 \cosh (4 \theta )-2)$ \\
        \hline
        \text{[12121212]} & $0$ & $\frac{7}{8} \sinh ^4(2 \theta )$ \\
        \hline
        \text{[12121222]} & $0$ & $\frac{7}{8} \sinh ^3(2 \theta ) \cosh (2 \theta )$ \\
        \hline
        \text{[12122122]} & $0$ & $\frac{1}{64} (7 \sinh (8 \theta )-6 \sinh (4 \theta ))$ \\
        \hline
        \text{[12122222]} & $0$ & $\frac{7}{32} \sinh ^2(4 \theta )$ \\
        \hline
        \text{[12212222]} & $\frac{1}{8}$ & $\frac{1}{16} \cosh ^2(2 \theta ) (7 \cosh (4 \theta )-5)$ \\
        \hline
        \text{[12221222]} & $0$ & $\frac{7}{32} \sinh ^2(4 \theta )$ \\
        \hline
        \text{[12222222]} & $0$ & $\frac{7}{8} \sinh (2 \theta ) \cosh ^3(2 \theta )$ \\
        \hline
        \text{[22222222]} & $\frac{7}{8}$ & $\frac{7}{8} \cosh ^4(2 \theta )$ \\
        \hline
    \end{NiceTabular}
    \caption{Expectation values of loops in $V_{8}$ at zero temperature and finite temperature (continued).}
    \label{tab:loop_info_V8_continued}
\end{center}
\end{table}

\clearpage
\newpage

\bibliographystyle{jhep}
\bibliography{reference}

\end{document}